\documentclass[preprint,prc,showpacs,preprintnumbers,
               superscriptaddress,amsmath,amssymb,floatfix]{revtex4}
\usepackage{graphicx}
\usepackage{dcolumn}
\usepackage{bm}
\usepackage{longtable}

\newcommand{\beq}{\begin{equation}}
\newcommand{\eeq}{\end{equation}}
\newcommand{\beqn}{\begin{eqnarray}}
\newcommand{\eeqn}{\end{eqnarray}}
\newcommand{\btab}{\begin{tabular}}
\newcommand{\etab}{\end{tabular}}

\newcommand{\brho}{\mbox{\boldmath$\rho$}}
\newcommand{\bome}{\mbox{\boldmath$\omega$}}

\usepackage{multirow,amsmath,booktabs}

\begin{document}

\title{Sensitivity of neutron radii in a $^{208}$Pb nucleus and a neutron star
to nucleon-sigma-rho coupling corrections in relativistic mean
field theory}

\author{G.~Shen}
\affiliation{School of Physics, Peking University, Beijing 100871}
\author{J.~Li}
\affiliation{School of Physics, Peking University, Beijing 100871}
\author{G.~C.~Hillhouse}
\thanks{e-mail: gch@sun.ac.za}
\affiliation{School of Physics, Peking University, Beijing 100871}
\affiliation{Department of Physics, University of Stellenbosch,
Stellenbosch, South Africa}
\author{J.~Meng}
\thanks{e-mail: mengj@pku.edu.cn}
\affiliation{School of Physics, Peking University, Beijing 100871}
\affiliation{Institute of Theoretical Physics, Chinese Academy of
Science, Beijing 100080}
\affiliation{Center of Theoretical Nuclear Physics, National Laboratory of \\
       Heavy Ion Accelerator, Lanzhou 730000}
\date{\today}

\begin{abstract}

We study the sensitivity of the neutron skin thickness, $S$, in a
$^{208}$Pb nucleus to the addition of nucleon-sigma-rho coupling
corrections to a selection (PK1, NL3, S271, Z271) of interactions
in relativistic mean field model. The PK1 and NL3 effective
interactions lead to a minimum value of $S$ = 0.16~fm in
comparison with the original value of $S$ = 0.28 fm. The S271 and
Z271 effective interactions yield even smaller values of $S$ =
0.11 fm, which are similar to those for nonrelativistic mean field
models. A precise measurement of the neutron radius, and therefore
$S$, in $^{208}$Pb will place an important constraint on both
relativistic and nonrelativistic mean field models. We also study
the correlation between the radius of a 1.4 solar-mass neutron
star and $S$.

\end{abstract}


\maketitle

\section{\label{sec:introduction}Introduction}
Precise values for proton and neutron densities in nuclei, as well
as their corresponding root-mean-square (rms) radii, are very
important for providing quantitative predictions in nuclear
physics and nuclear astrophysics. Our present understanding of
nuclear phenomena largely hinges on accurate and model independent
determinations of charge densities and radii via electron
scattering. However, this picture of the nucleus will be more
complete once precise values for neutron radii become available.
Currently, the situation regarding the experimental determination
of neutron radii is unsatisfactory with errors typically being an
order of magnitude larger than those for proton radii
\cite{Fortson}. What is more disconcerting, however, is the large
variation in quoted values of the neutron radius for a single
nucleus such as $^{208}$Pb \cite{Trz,Allar,staro,kras,kara,clark}.
This dismal situation has prompted an experiment at Jefferson
Laboratory \cite{Jeff} to measure the neutron radius in $^{208}$Pb
accurately and model independently via parity-violating electron
scattering, the results of which are expected to have a widespread
and important impact in nuclear physics.

Reliable neutron densities and radii are needed for quantitatively
characterizing the bulk properties of nuclear matter at normal nuclear
densities, which in turn serve as valuable input for: studying the dynamics
of heavy-ion collisions, determining the equation-of-state which
underpins the evolution and structure of supernovae and neutron
stars, studying the properties of exotic neutron-rich nuclei far
from the beta-stability line, reducing uncertainties in atomic
parity violation experiments, as well as for studying pionic and
antiprotonic atoms.

Recently, much attention
\cite{Trz,Allar,staro,kras,kara,clark,Typel,Brown,Furn} has been
devoted to studying the neutron skin thickness, $S$, in
$^{208}$Pb, a quantity which is defined as the difference between
the rms radius of neutrons, $\sqrt{<r_n^2>}$, and protons,
$\sqrt{<r_p^2>}$, {\it i.e.}, $S$ = $\sqrt{<r_n^2>} -
\sqrt{<r_p^2>}$. A precise measurement of the neutron radius of
$^{208}$Pb, together with existing high-precision measurements of
the proton radius \cite{Fricke}, will yield a precise value for
the neutron skin thickness, thus providing one of the most stringent tests
to date for current models of nuclear structure. Indeed, the
Parity Radius Experiment at Jefferson Laboratory aims to measure
the neutron radius in $^{208}$Pb to an unprecedented accuracy of
1\% ($\pm$0.05~fm). Currently relativistic and nonrelativistic
models predict completely different values for the neutron radius
and, hence, it is evident that the latter measurement will have an
impact on deepening our understanding of the dynamical basis of
nuclear structure. In particular, nonrelativistic Skyrme models
predict $S$ = 0.1 $\sim$ 0.2~fm \cite{Typel,Brown,Furn}, whereas
present relativistic mean field (RMF) models, on the other hand, give $S$
= 0.2 $\sim$ 0.3~fm \cite{Typel,Furn}. Model-dependent analyses of
experimental data with hadronic probes yield values of $S$ varying
between 0.0 and 0.2~fm (see Table \ref{tab:data}), which seem to
be more consistent with the predictions of nonrelativistic models
based on the Schr\"{o}dinger equation. Hadronic measurements,
however, suffer from potentially serious theoretical systematic
errors associated with uncertainties in nuclear reaction
mechanisms and, hence, one should be cautious about drawing
conclusions regarding the appropriateness of various dynamical
models.

In this paper we focus on a relativistic description of nuclei and
neutron stars within the field theoretical framework of quantum
hydrodynamics. In particular, we capitalize on the success of the
Walecka model and extensions thereof to successfully describe the
properties of nuclear matter and finite nuclei \cite{meng98npa,
meng98prl,MTY98,MZT02,MTZZZ02}, as well as predict the well-known
central and spin-orbit potentials which are usually postulated in
nonrelativistic Schr\"{o}dinger-equation-based models
\cite{Gino,MSY98,MT99,MSY99}. More specifically, we supplement
existing relativistic mean field Lagrangian densities with two new
isospin dependent higher order correction terms relating to the
coupling of the nucleon current to sigma- and rho-meson fields.
These terms can be associated with multi-meson exchange processes
occurring in the inner higher-density region of nuclei. In
low-energy nucleon-nucleon scattering models, the contribution to
the scattering amplitudes is usually dominated by single-meson
exchange with an effective mass of less than 1 GeV. However, it
has been demonstrated that the inclusion of two-meson exchange at
low energies provides a major improvement in the description of
nucleon-nucleon scattering data \cite{Th}. In general there is no
unique way for introducing nonlinear isovector-nucleon couplings
in relativistic mean field models. In this paper we introduce the
simplest structure generating the nonlinear isospin-dependent
terms. Values for the various combinations of the new coupling
constants are extracted by fitting to the properties of nuclear
matter -- such as saturation density, binding energy per nucleon,
nuclear incompressibility and the symmetry energy -- and the
corresponding values for the neutron skin thickness in $^{208}$Pb
are extracted for a selection of RMF models. In particular, we
focus on the PK1 \cite{Long}, NL3 \cite{NL3}, S271 and Z271
\cite{H1} RMF parameterizations.

One of the issues we wish to address in the paper is to find ways
of adjusting the neutron radius in $^{208}$Pb by adding new terms
to the Lagrangian density of existing relativistic mean field
models, while at the same time keeping the value of proton radius
(which is accurately determined experimentally) fixed. In
particular, the question arises as to whether it is possible for
RMF models to produce values of the neutron radius which are in
the same range as those for nonrelativistic models.

In recent analyses, a linear relationship between the neutron skin
in $^{208}$Pb and the nuclear symmetry energy at saturation
density was proposed \cite{Furn,Brown,Diep}. We also investigate
the relationship between the latter quantities for our new RMF
models. The new interaction terms considered in this paper
influence the density dependence of
symmetry energy, which in turn affects the neutron skin in
$^{208}$Pb. In addition, we extrapolate from normal to dense
neutron matter and study the correlation between the radius of a
neutron star and the neutron radius in $^{208}$Pb. Indeed,
Horowitz and Piekarewicz \cite{H3} have performed such a
correlation analysis for a wide range of RMF models and they
concluded that, whereas the radius of a 0.5 solar-mass neutron
star can be inferred from the radius of $^{208}$Pb, the radius of
a 1.4 solar-mass neutron star is not uniquely constrained by a
measurement of the neutron skin thickness. In this paper we
perform a similar analysis for the PK1, NL3, S271 and Z271
effective interactions supplemented with our new terms, but for
simplicity we only consider the mass and radius of a 1.4
solar-mass neutron star.

This paper is organized in the following sequence. A brief
formulation of our new RMF model in relation to existing RMF
models for finite nuclei and neutron star matter is presented in
Sec.~\ref{sec:formalism}. The extraction of parameter sets for our
new RMF models is discussed in Sec.~\ref{sec:extraction}. In
Sec.~\ref{sec:results} results are presented and discussed for the
neutron radius in $^{208}$Pb. Values of neutron radii in
$^{208}$Pb for different parameter sets are then correlated with
the radius of a 1.4 solar-mass neutron star. Finally, in
Sec.~\ref{sec:summary}, we summarize the main points of this
paper.

\section{\label{sec:formalism}Higher order corrections to RMF models for spherical nuclei and neutron stars}

The basic physics underlying RMF models and their application in
nuclear physics can be found in Refs.~\cite{Serot,Rein,Ring}. In
this section we present our new RMF model and indicate its
relation to existing RMF models. The basic ansatz of the RMF
theory is a Lagrangian density whereby nucleons are described as
Dirac particles which interact via the exchange of sigma-
($\sigma$), omega- ($\omega$), and rho- ($\rho$) mesons, and also
the photons ($A$), namely: \beqn\label{lagrangian}
    {\cal L}&=&\overline{\psi}\left[i{\gamma^\mu}
              {\partial_\mu}-m-{g_\sigma}\sigma - g_\omega
              \gamma^\mu\omega_\mu - g_\rho \gamma^\mu \vec{{\bf
              \tau}}\cdot \vec{\mbox{\boldmath$\rho$}}_\mu - e\gamma^\mu\frac
              {1+\tau_3}{2} A_\mu\right]\psi\nonumber\\
            &&+\ \frac{1}{2}\partial^\mu\sigma\partial_\mu\sigma-\frac{1}{2}m_\sigma^2\sigma^2-
              U(\sigma)\nonumber\\
            &&-\
              \frac{1}{4}\omega^{\mu\nu}\omega_{\mu\nu}+\frac{1}{2}m_\omega^2\omega^\mu\omega_\mu+
              U(\omega)\nonumber\\
            &&-\ \frac{1}{4}\vec{\mbox{\boldmath$\rho$}}^{\mu\nu}
              \cdot\vec{\mbox{\boldmath$\rho$}}_{\mu\nu}+
              \frac{1}{2}m_\rho^2\vec{\mbox{\boldmath$\rho$}}^\mu
              \cdot\vec{\mbox{\boldmath$\rho$}}_\mu\nonumber\\
            &&-\ \frac{1}{4}A^{\mu\nu}A_{\mu\nu}
\eeqn where the field tensors of the vector mesons and the
electromagnetic field take the following form:
\beqn\label{tensor}
   \omega^{\mu\nu}&=&\partial^\mu\omega^\nu-\partial^\nu\omega^\mu,\nonumber\\
                     A^{\mu\nu}&=&\partial^\mu A_\nu -\partial^\nu A_\mu,\nonumber\\
                     \vec{\mbox{\boldmath$\rho$}}^{\mu\nu}&=&\partial^\mu\vec{\mbox{\boldmath$\rho$}}^\nu-
                     \partial^\nu\vec{\mbox{\boldmath$\rho$}}^\mu-g_\rho\vec{\mbox{\boldmath$\rho$}}^\mu\times
                     \vec{\mbox{\boldmath$\rho$}}^\nu\,.
\eeqn The nonlinear self-couplings for $\sigma$ and $\omega$
mesons are respectively:
\beqn\label{nonlinear}
   \begin{array}{cc} U(\sigma)\ =\ \frac{1}{3}g_2\sigma^3\ +\
                     \frac{1}{4}g_3\sigma^4,& U(\omega)\ =\
                     \frac{1}{4}c_3\left(\omega^\mu\omega_\mu\right)^2,\\
   \end{array}
\eeqn
with the self-coupling constants $g_2$, $g_3$, and $c_3$.

To modify the density dependence of nuclear matter symmetry
energy, Horowitz and Piekarewicz \cite{H1} introduced the
following nonlinear omega-rho coupling term:
\beqn\label{LHP}
    {\cal L}_{HP} =\ 4\Lambda_v g_\rho^2 \vec{\rho_{\mu}}\cdot \vec{\rho^\mu}
                     g_{\omega}^2\omega_\mu \omega^\mu\,.
\eeqn

In addition to considering the above term, in this paper we
introduce, for the first time, the following two new couplings of
the nucleon current to sigma- and rho-meson fields:
\beqn\label{L1}
    {\cal L}_1 =\ -\Gamma_1 \bar{\psi}g_\rho\gamma^\mu
               (g_{\sigma}\sigma / m)\ \vec{\tau}\cdot \vec{\rho_{\mu}}\psi
\eeqn
\beqn\label{L2}
    {\cal L}_2 =\ -\Gamma_2 \bar{\psi}g_\rho\gamma^\mu
               (g_\sigma \sigma / m )^2 \vec{\tau}\cdot \vec{\rho_{\mu}}\psi
\eeqn
{\it i.e.}, the total Lagrangian density of interest is:
\beqn\label{total}
      {\cal L}' =\ {\cal L}\ + {\cal L}_{HP}\ +\
            {\cal L}_1\ +\ {\cal L}_2\,.
\eeqn

The classical variational principle leads to the following
equations of motion:
\beqn\label{nucleon-motion}
     [\mathbf{\sl{\alpha}}\cdot\mathbf{\sl{p}}\ +\ V(\mathbf{r})\ +\
     \sl{\beta} (m\ +\ S(\mathbf{r}))] \psi_i\ =\ \varepsilon_i\psi_i
\eeqn
for the nucleon spinors, where
\beqn \label{Dirac}
     \left \{
     \begin{array}
        {l} V(\mathbf{r}) =\ \beta \{g_{\omega}\rlap{/}{\omega}_{\mu}\
        +\ {g_{\rho}} (1\ +\ \Gamma_1 (g_{\sigma}\sigma / m)\ +\ \Gamma_2
        (g_\sigma \sigma / m )^2 )\vec{\tau}\cdot
        \rlap{/}\vec{\rho_{\mu}}\  +\ e \frac{ (1\ +\ \tau_3)}{2}\ \rlap{/}A_{\mu} \}\\
        S(\mathbf{r}) =\ g_{\sigma}\sigma\
     \end{array}
     \right.
\eeqn

and
\beqn\label{K-G}
    \left \{
    \begin{array}{l}
       (m_{\sigma}^2\ -\ \nabla^2)\
       \sigma =\ -g_\sigma \rho_s -\ g_2 \sigma^2\ -\ g_3\sigma^3 -
       {g_{\rho}} \vec{\rho}_{\mu} \cdot\vec{j}^\mu(\Gamma_1
       (g_{\sigma}\sigma / m)\ +\ 2 \Gamma_2 ( g_\sigma^2 \sigma / m ^2))\\
       (m_v^2\ -\ \nabla^2)\omega^\mu =\ g_\omega j^\mu -c_3\omega^\mu(\omega^\mu\omega_\mu)-8\Lambda_v
       g_{\rho}^2g_\omega^2 \vec{\rho}_{\mu} \cdot \vec{\rho}^{\mu}\omega^\mu \\
       (m_{\rho}^2\ -\ \nabla^2) \vec{\rho}^{\mu} =\ {g_{\rho}}
       \vec{j}^\mu(1\ +\ \Gamma_1 (g_{\sigma}\sigma / m)\ +\ \Gamma_2 (
       g_\sigma \sigma / m )^2 ) -8\Lambda_v
       g_{\rho}^2g_\omega^2\omega_\mu \omega^\mu \vec{\rho}^{\mu}  \\
       \label{speqa}\hspace{2.5em} -\ {\nabla}^2A^\mu =\ ej^\mu_{p}
    \end{array}
    \right.
\eeqn for the mesons and photons. The nucleon spinors provide the
relevant source terms:
\beqn\label{source}
    \left \{
    \begin{array}{l}
      \rho_s =\ \sum_{i = 1}^A\overline{\psi}_i\psi_i \\
      j^\mu =\ \sum_{i = 1}^A\overline{\psi}_i \gamma^\mu \psi_i \nonumber\\
      \vec{j}^\mu =\ \sum_{i = 1}^A \overline{\psi}_i \gamma^\mu
      \vec{\tau} \psi_i\\
      \label{speqb} j^\mu_{p} =\ \sum_{i =
      1}^A\overline{\psi}_i\gamma^\mu\frac {1+\tau_3}{2}\psi_i
    \end{array}
    \right.
\eeqn where the summations run over the valence nucleons only. The
present method neglects the contribution of negative energy
states, {\it i.e.} the so-called no sea approximation. The above
non-linear equations are solved by iteration within the context of
the mean field approximation whereby the meson field operators are
replaced by their expectation values.

\subsection{Spherical nuclei}

Since spherical nuclei such as $^{208}$Pb respect time reversal
symmetry, there are no currents in the nucleus and the spatial
vector components $\bome$, $\brho$ and {\bf $A$} vanish. One is
left with the time-like components ${\omega}_0$, $\vec{\rho}_0$
and $A_0$. Charge conservation guarantees that only the
3-components of the isovector $\rho_{0,3}$ survive. As the system
obeys rotational symmetry, the nucleon potentials and the
meson-field sources depend only on the radial coordinate $r$. The
Dirac spinor, $\psi (r)$ is characterized by the angular momentum
quantum numbers $l, j, m$, the isospin $t = + (-)$ for protons
(neutrons) respectively, and any other relevant quantum numbers
$i$, i.e. \beqn\label{wavefunction}
    \psi (r) =\ \left(
                \begin{array}{c}
                     i\frac{G^{lj}_i(r)}{r}Y^l_{jm}(\theta,\phi)\\
                     \frac{F^{lj}_i(r)}{r}(\vec{\sigma}\cdot\hat{r})Y^l_{jm}(\theta,\phi)
                \end{array}
                \right)\chi_t
\eeqn where $Y^l_{jm}(\theta,\phi)$ are the usual spinor spherical
harmonics, $G^{lj}_i(r)/r$ and $F^{lj}_i(r)/r$ denote the upper-
and lower-component radial wave functions, and $\chi_t$ represents
the isospin wave functions specified by the isospin $t$. The
radial wave functions are normalized as:

\beqn
     \int^\infty_0 dr (\mid G^{lj}_i(r)\mid^2\ +\
     \mid F^{lj}_i(r)\mid^2) = 1\,.
\eeqn

The following phase convention is adopted for the vector spherical harmonics:
\begin{equation*}
      (\vec{\sigma}\cdot\hat{r})Y_{jm}^l\ =\ -Y_{jm}^{l'},
\end{equation*}
where
\begin{equation*}
      l' = 2j-l = \left\{
       \begin{array}{cc} l+1, & j = l + 1/2 , \\
                          l-1 ,   & j = l - 1/2.
       \end{array} \right.
\end{equation*}

After some tedious algebra, one gets the radial equation for the
upper and lower components, respectively \cite{meng98npa}:
\beqn
\left \{ \begin{array}{l} \epsilon_i G^{lj}_i(r) = \left (-\frac
\partial {\partial r}\ +\ \frac{\kappa_i}{r}\right)F^{lj}_i(r)\
+\ (m\ +\ S(r)\ +\ V(r))G^{lj}_i(r) \\
 \epsilon_i F^{lj}_i(r) =
\left(+\frac \partial {\partial r}\ +\
\frac{\kappa_i}{r}\right)G^{lj}_i(r)\ -\ (m\ +\ S(r)\ -\
V(r))F^{lj}_i(r)
\end{array}
\right. \eeqn
where
\begin{equation*}
      \kappa = \left\{
       \begin{array}{cc}
             -(j+1/2), & \mathrm{for}\ j = l + 1/2,  \\
             +(j+1/2) ,   &\mathrm{for}\  j = l - 1/2,
       \end{array}
      \right.
\end{equation*}
and
\beqn
     \left \{
      \begin{array}{l}
         V(r) =\ g_{\omega}{\omega_{0}\
         +\ {g_{\rho}} (1\ +\ \Gamma_1 (g_{\sigma}\sigma / m)\ +\ \Gamma_2
         (g_\sigma \sigma / m )^2 )\tau_3 \rho_{0,3}}\  +\ \frac{1}{2} e (1\ +\ \tau_3) A_{0},\\
         S(r) =\ g_{\sigma}\sigma .
      \end{array}
     \right.
\eeqn

The meson fields equations reduce to radial Laplace equations of the
form:
\beqn
     \left( \frac{\partial^2}{\partial r^2}\ -\ \frac 2 r
     \frac {\partial}{\partial r}\ +\ m_\phi^2 \right)\phi\ =\ S_\phi
\eeqn where $m_\phi$ are the meson masses for $\phi\ =\ \sigma,\
\omega,\ \rho$ and $m_\phi\ =\ 0$ for the photon. The relevant
source terms are:
\beqn
     S_\phi = \left\{
       \begin{array}{cc}
                 -g_\sigma \rho_s -\ g_2 \sigma^2\ -\ g_3\sigma^3 - {g_{\rho}}
                 \rho_{0,3}\rho_3 (\Gamma_1 (g_{\sigma}\sigma / m)\ +\ 2 \Gamma_2
                 ( g_\sigma^2 \sigma / m ^2)) &\mathrm{for\ the\ \sigma\ field},  \\
                 g_\omega\rho_\upsilon\ -\ c_3\omega^3_0 -\ 8\Lambda_v g_{\rho}^2g_\omega^2 \rho_{0,3}^2 \omega_0
                  &\mathrm{for\ the\ \omega\ field},\\
                 {g_{\rho}} \rho_3 (1\ +\ \Gamma_1 (g_{\sigma}\sigma / m)\ +\
                 \Gamma_2 ( g_\sigma \sigma / m )^2 ) -8\Lambda_v
                 g_{\rho}^2g_\omega^2\omega_0^2 \rho_{0,3} & \mathrm{for\ the\ \rho\ field},\\
                 e \rho_c(r) & \mathrm{for\ the\ Coulomb\ field},
       \end{array}
     \right.
\eeqn
where the various nucleon densities are given by:
\beqn
      \left \{
      \begin{array}{l}
         4\pi r^2 \rho_s(r) =\ \sum_{i=1}^A (|G_i(r)|^2\ -\ |F_i(r)|^2),\\
         4\pi r^2 \rho_\upsilon(r) =\ \sum_{i=1}^A (|G_i(r)|^2\ +\ |F_i(r)|^2), \\
         4\pi r^2 \rho_3(r) =\ \sum_{i=1}^Z(|G_i(r)|^2\ +\ |F_i(r)|^2)\ -\
          \sum_{i=1}^N(|G_i(r)|^2\ +\ |F_i(r)|^2), \\
         \label{speq}4\pi r^2 \rho_c(r) =\ \sum_{i=1}^Z(|G_i(r)|^2\ +\ |F_i(r)|^2).
      \end{array}
      \right.
\eeqn The neutron and proton rms radii are directly related to the
neutron and proton density distributions via the following
relationship: \beqn\label{rms}
     R_i\ =\ \sqrt{<r_i^2>}\ =\ \sqrt{\frac{\int  \rho_i(r) r^2d^3r}{\int
     \rho_i(r)d^3r}}
\eeqn where $\rho_i$(r) ($i = n, p$) denotes the corresponding
neutron and proton baryon density distributions.

The total binding energy of the system is: \beqn\label{energy} E
&=& E_{nucleon} + E_\sigma + E_\rho + E_\omega + E_c - mA \cr &=&
\sum_i \epsilon_i\ -\ \frac 1 2 \int d^3r \{g_\sigma \sigma
\rho_s(r)\ +\ \frac 1 3 g_2\sigma^3\ +\ \frac 1 2 g_3\sigma^4\}\
\cr && -\ \frac 1 2 \int d^3rg_\rho\rho_{0,3}\{(1\ +\ 2 \Gamma_1
\frac{g_\sigma\sigma}{m}\ +\ 3
\Gamma_2(\frac{g_\sigma\sigma}{m})^2)\rho_3(r)\ -\ 8 \Lambda_v
(g_\omega \omega_0 g_\rho\rho_{0,3} )^2\} \cr && -\ \frac 1 2 \int
d^3r\{g_\omega \omega_0\ -\ \frac 1 2 c_3\omega_0^4\}\rho_v(r) -\
\frac 1 2 \int d^3r A_0 \rho_c(r) - mA\, . \eeqn

Eq.~(13) is solved self-consistently using the Runge-Kutta
algorithm and the shooting method, and the Laplace equations,
given by Eq.~(15), are solved employing a Green Function method.

\subsection{Neutron stars}
Next we discuss the RMF Lagrangian density used for describing a
neutron star consisting of only nucleons and leptons (mainly
electrons, $e^-$, and muons, $\mu^-$). The Lagrangian density associated with the
leptons is: \beqn\label{leptonic}
     {\cal L}_{lepton} &=&
      \sum_{\lambda=e^-,\mu^-}{\bar\psi}_\lambda(i\gamma^\mu\partial_\mu
     - m_\lambda)\psi_\lambda ,
\eeqn such that the full Lagrangian density under consideration
consists of a hadronic part, given by Eq.~(\ref{total}), plus a
leptonic part, given by Eq.~(\ref{leptonic}). Introducing the mean
field and no-sea approximations, the equations of motion for
baryons and mesons can be derived. The detailed formalism of which
can be found in Ref.~\cite{Ban}. The Dirac equation represents the
equation of motion associated with the leptons. Based on chemical
equilibrium and charge neutral conditions, for a given total
baryon density, the equations of motion are solved
self-consistently by iteration, and the meson fields, particle
densities and Fermi momenta of each species (protons, neutrons,
electrons, and muons) are obtained simultaneously.

The canonical energy-momentum tensor is derived by invoking
the invariance of spacetime translation, namely: \beqn\label{e-p-1}
   {\cal T^{\mu\nu}} &=& -g^{\mu\nu}{\cal L}\ +\ \sum_\phi
   \frac{\partial{\cal
   L}}{\partial(\partial_\mu\phi)}\partial^\nu\phi,
\eeqn where $g^{\mu\nu}$ is the Minkowski metric in rectilinear
coordinates. For the case in which neutron star matter can be
considered to be a perfect fluid, the energy-momentum tensor is
given by \cite{GL}:
\beqn\label{e-p-2} {\cal T^{\mu\nu}} &=&
   \left(
  \begin{array}{cccc}
    \epsilon& 0&0&0\\
    0&p&0&0\\
    0&0&p&0\\
    0&0&0&p
    \end{array}
   \right).
\eeqn Comparing the ground-state expectation value of
Eq.~(\ref{e-p-1}) with Eq.~(\ref{e-p-2}), via the field equations
the energy density, $\epsilon$, and the pressure, $P$, are given by: \beqn\label{e-p}
     \epsilon &=&\frac{1}{2}m_\sigma^2\sigma^2 + \frac{1}{3}g_2\sigma^3 + \frac{1}{4}g_3\sigma^4
        + \frac{1}{2}m_\omega^2\omega_0^2 + \frac{3}{4}c_3\omega_0^4
        + \frac{1}{2}m_\rho^2\rho_{0,3}^2
        + 12g_\rho^2g_\omega^2\Lambda_v\omega_0^2\rho_{0,3}^2\nonumber\\
&&
    + \frac{1}{\pi^2} \sum_{B=n,p}\int_0^{k_B}k^2dk\sqrt{k^2 + (m_B + g_\sigma\sigma)^2}
    +\ \frac{1}{\pi^2}\sum_{\lambda=e^-,\mu^-}\int_0^{k_\lambda}k^2dk
\sqrt{k^2\ +\ m_\lambda^2},
\\
P &=&
    -\frac{1}{2}m_\sigma^2\sigma^2 - \frac{1}{3}g_2\sigma^3 - \frac{1}{4}g_3\sigma^4
    + \frac{1}{2}m_\omega^2\omega_0^2 + \frac{1}{4}c_3\omega_0^4
    + \frac{1}{2}m_\rho^2\rho_{0,3}^2
    + 4g_\rho^2g_\omega^2\Lambda_v\omega_0^2\rho_{0,3}^2
    \nonumber\\
&&
    + \frac{1}{3\pi^2}\sum_{B=n,p}\int_0^{k_B}\frac{k^4}{\sqrt{k^2 + (m_B + g_\sigma\sigma)^2}}dk
    + \frac{1}{3\pi^2}\sum_{\lambda=e^-,\mu^-}\int_0^{k_\lambda}dk
\frac{k^4}{\sqrt{k^2 + m_\lambda^2}}\, ,
 \eeqn where $k_B$ and $k_\lambda$ are the Fermi momenta of
 baryons and leptons respectively.

The mass and radius of a neutron star are obtained by employing
the Oppenheimer-Volkoff (OV) equations
\cite{oppenheimer39,tolman39}:
\beqn\label{O-V}
     &&\frac{dp(r)}{dr} =
    - \frac{[p(r) + \epsilon(r)][M(r) + 4\pi r^3p(r)]}{r(r - 2M(r))}\, ,\\
     &&M(r) = 4\pi\int_0^r\epsilon(r^\prime)r^{\prime2}dr^\prime\, ,
\eeqn where $r$ denotes the radial coordinate relative to the
center of the star, $p(r)$ and $\epsilon(r)$ are the pressure and
energy density at a radial point $r$ in the star respectively,
and $M(r)$ represents the mass of the sphere contained within a radius $r$.
Since zero pressure cannot support a neutron star from collapsing,
we define the radius, $R$, of the star as that radius at which the
pressure is zero. The mass total mass, $M(R)$, of the star is
subsequently defined as the mass contained within a sphere of
radius $R$.

\section{\label{sec:extraction}Extraction
of parameter sets for new RMF models}

In this section we present the procedure and criteria for extracting
values for our new coupling constants $\Gamma_{1}$ and $\Gamma_{2}$
in Eqs.~(\ref{L1}) and (\ref{L2}) respectively. In particular, we consider the
addition of these new isopsin-dependent higher order correction
terms to the PK1 \cite{Long}, NL3 \cite{NL3}, S271 and Z271 \cite{H1}
RMF models. The parameters sets for these interactions
are presented in Table \ref{tab:para}.

We start with the recent PK1 effective interaction \cite{Long}.
The PK1 effective interaction provides an excellent description of
the properties of nuclear matter as well as for nuclei near and
far from the valley of $\beta$ stability. For this interaction,
symmetric nuclear matter saturates at a Fermi momentum of
1.30~fm$^{-1}$ with a binding energy per nucleon of $-$16.27~MeV
and an incompressibility of $K$ = 283~MeV. For comparison with the
results of the PK1 effective interaction, we also employ the NL3,
S271, and Z271 effective interactions, which have been used in
Refs.~\cite{H1,H3} to observe the effects of adding a nonlinear
omega-rho coupling term, given by Eq.~(\ref{LHP}), on the neutron
skin thickness, $S$, in $^{208}$Pb and the radius of 1.4
solar-mass neutron star. The NL3 effective interaction has also
been used extensively to reproduce a variety of nuclear
properties, such as binding energies, nuclear radii, nuclear
density distribution, single particle spectra, etc. The NL3 and
S271 effective interactions contain a sigma-meson self-coupling
and Z271 includes both sigma- and omega-meson self-couplings.
These three interactions produce the following properties for
symmetric nuclear matter: saturation at a Fermi momentum of
1.30~fm$^{-1}$, a binding energy per nucleon of $-$16.24~MeV, and
an incompressibility of $K$ = 271~MeV.

Later in this section we will dicuss the procedure for calibrating
values of the $\Gamma_{1}$ and $\Gamma_{2}$ couplings, in
Eqs.~(\ref{L1}) and (\ref{L2}), with respect to the symmetry
energy, a quantity which we now briefly discuss. The energy of
asymmetric nuclear matter can be expanded around the energy of
symmetric nuclear matter -- where symmetric nuclear matter is
characterized by identical proton and neutron densities,
$\rho_p$ and $\rho_n$ respectively -- as: \beqn\label{asymmetric}
     \frac E A (\rho,t)\ =\ \frac E A (\rho,t=0)\ +\ t^2 a_{sym}(\rho)\ +\
     O(t^4),
\eeqn where $\rho = \rho_n + \rho_p = \frac {2k_F^3} {3\pi^2}$,
$k_F$ is the the Fermi momentum, $a_{sym}(\rho)$ is the symmetry
energy, and the asymmetry, $t$, is defined as: \beqn
  t \equiv \frac{\rho_n-\rho_p}{\rho_n+\rho_p}\ .
\eeqn

The symmetry energy, $a_{sym}(\rho)$, describes how the energy of
asymmetric nuclear matter changes with the asymmetry, $t$.
>From Eq.~(27), and employing Eq.~(19), one can extract the following
expression for the symmetry energy of nuclear matter in RMF:
 \begin{equation}
    \label{symmetry}
       a_{sym}(\rho)\ =\ (\frac{\partial^2 \frac E A(\rho,t)}{\partial
       t^2})_{t=0}\ =\ \frac {k_F^2}{6E_F^*}\ +\ \frac{{g_\rho}^2}{12\pi^2}\
                \frac{k_F^3}{{m^*_\rho}^2},
 \end{equation}
where $E_F^{*2} = k_F^2 + m^{*2}$, $m^{*} = m + g_{\sigma}\sigma$
is the effective nucleon mass, and the effective rho-meson mass,
$m^*_\rho$, is defined as: \beqn\label{effectiverho}
     {m^*_\rho}^2\ =\ \frac{m_\rho^{2}+
     8\Lambda_vg_\rho^2g_{\omega}^2\omega_0^2}{(1\ +\ \Gamma_1
     \frac{g_{\sigma}\sigma}{m} +\
     \Gamma_2(\frac{g_{\sigma}\sigma}{m})^2)^2}\,.
\eeqn

There are two terms in the symmetry energy. The first term in
Eq.~(\ref{symmetry}) represents the increase in the kinetic energy of
the system when there is a relative displacement of the neutron and proton
Fermi energies, and the second term in Eq.~(\ref{symmetry}) represents the
coupling between effective rho-meson and an isovector-vector current
which no longer vanishes in asymmetric nuclear matter.

We now describe our procedure for extracting values of $g_{\rho}$,
$\Lambda_{v}$, and $\Gamma_{1}$, with $\Gamma_{2} = 0$, in the
Lagrangian density given by Eq.~(\ref{total}). The symmetry energy
at saturation density, corresponding to $k_f$ = 1.30~fm$^{-1}$, is
not well constrained by the binding energy of nuclei. However,
some average of the symmetry energy at full density and the
surface energy is constrained by the binding energy \cite{H1}. We
adopt the procedure of Horowitz and Piekarewicz, whereby we choose
values of $\Lambda_{v}$ and $\Gamma_{1}$ with $\Gamma_{2}$ = 0,
and then adjust the value $g_{\rho}$ such that the effective
interaction has a fixed symmetry energy at an average density of
$k_f$ = 1.15~fm$^{-1}$ ($\rho$ = 0.10~fm$^{-3}$) and the binding
energy per nucleon in $^{208}$Pb lies in the range $| E/A -
(E/A)_{exp} |\le 0.005$ MeV, where $(E/A)_{exp}$ represents the
experimental value. The symmetry energy at $k_f$ = 1.15~fm$^{-1}$
is 26.08~MeV for the original PK1 effective interaction and
25.68~MeV for the original NL3, S271, and Z271 effective
interactions, respectively. A similar procedure is followed for
extracting parameter values for $g_{\rho}$, $\Lambda_{v}$, and
$\Gamma_{2}$, with $\Gamma_{1} = 0$.

The experimental binding energy per nucleon, $(E/A)_{exp}$, in
$^{208}$Pb is $-$7.868~MeV \cite{Audi}. As the correction for
center of mass motion is neglected in the self-consistent RMF
calculation, for comparative purposes an ``equivalent''
experimental $(E/A)_{exp}$ = $-$7.843~MeV for $^{208}$Pb is used
instead, which is estimated via the harmonic oscillator phenomenological
formula $E_{cm}\ =\ 0.75\ \hbar\omega_0$ with $\hbar\omega_0 = 41 A^{-1/3}$
\cite{meng98npa}.

\begin{figure}[htbp]
 \centering
 \includegraphics[height=15cm,angle=-90]{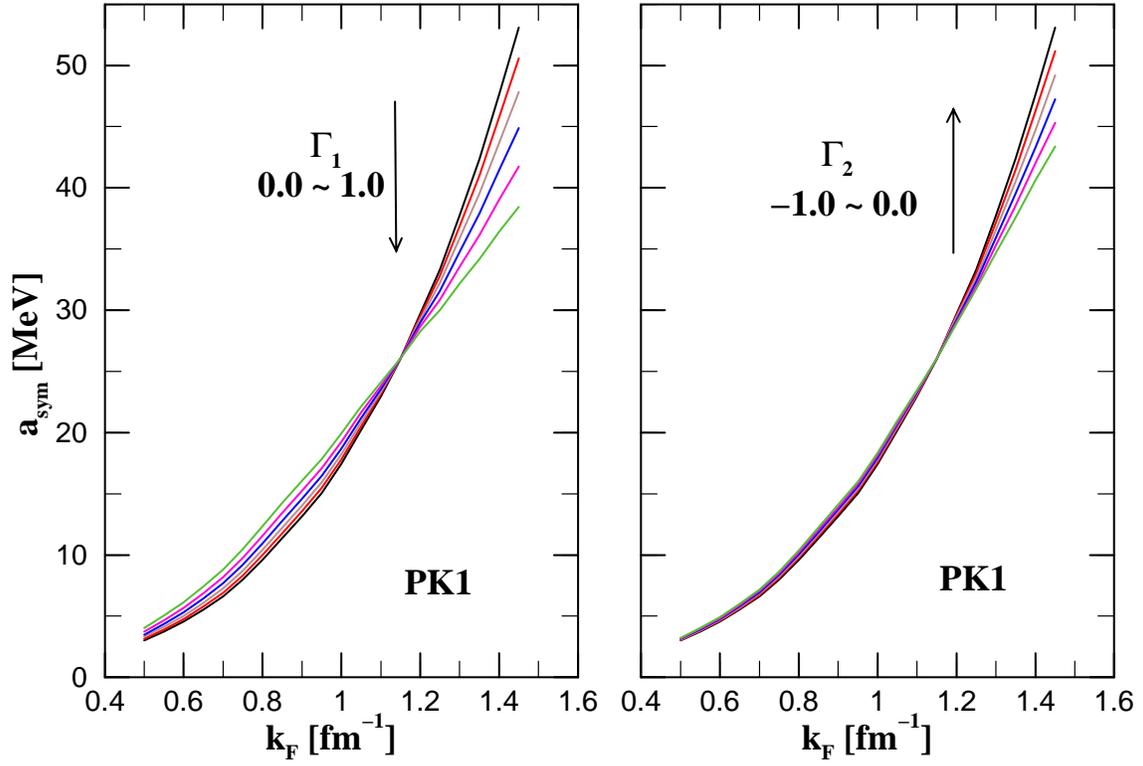}
 \caption{(Color online) The symmetry energy, $a_{sym}$ (in MeV), in symmetric
nuclear matter as a function of the Fermi momentum, $k_{F}$ (in
fm$^{-1}$), after adding different values of $\Gamma_1$ (left
panel) or $\Gamma_2$ (right panel) terms to the original PK1
effective interaction. The direction of the arrows indicates values
of increasing $\Gamma_{1}$ and $\Gamma_{2}$.}
 \label{fig:pk1-g1+g2}
\end{figure}

To avoid abnormal solutions at zero density for nuclear matter,
$\Lambda_v$ in Eq.~(\ref{LHP}) should not be negative. However,
there is no such constraint on the values of $\Gamma_1$ and
$\Gamma_2$ in Eqs.~(5) and (6) respectively.  The role of the
$\Gamma_{1}$, $\Gamma_{2}$ and $\Lambda_{v}$ couplings is
to modify the density dependence of the symmetry energy.
In Fig.~\ref{fig:pk1-g1+g2}, we see that increasing $\Gamma_1$,
decreasing $\Gamma_2$, or increasing $\Lambda_v$ (see Ref.~\cite{H3}),
causes the symmetry energy to grow more slowly with density.

After readjusting the coupling constant g$_\rho$, we perform
self-consistent calculations as stated in Sec.~\ref{sec:formalism}
to determine the properties of $^{208}$Pb and a 1.4 solar-mass neutron
star. The numerical values of different combinations of $\Lambda_v$,
$\Gamma_1$ (or $\Gamma_2$) and $g_\rho$, and the corresponding binding
energy per nucleon, $E/A$, the proton root mean square radius,
$R_p$, the neutron skin thickness, $S$, and the radius, $R$, of a 1.4
solar-mass neutron star, are also listed in Tables~\ref{tab:PK11} to
\ref{tab:Z2711}.

\section{\label{sec:results}Results and Discussion}

\subsection{Neutron radius of $^{208}$Pb}

In this section we study the sensitivity of the neutron skin
thickness, $S$, in $^{208}$Pb with respect to the addition of two
new terms, given by Eqs~(5) and (6), to the original Lagrangian
densities for each of the PK1, NL3, S271 and Z271 effective
interactions. In particular we also extract values of the neutron
rms radius for RMF parameter sets which are consistent with
experimental binding energies as well as with the properties of
nuclear matter specified in the previous section.

\begin{figure}[htbp]
 \centering
 \includegraphics[height=15cm,angle=-90]{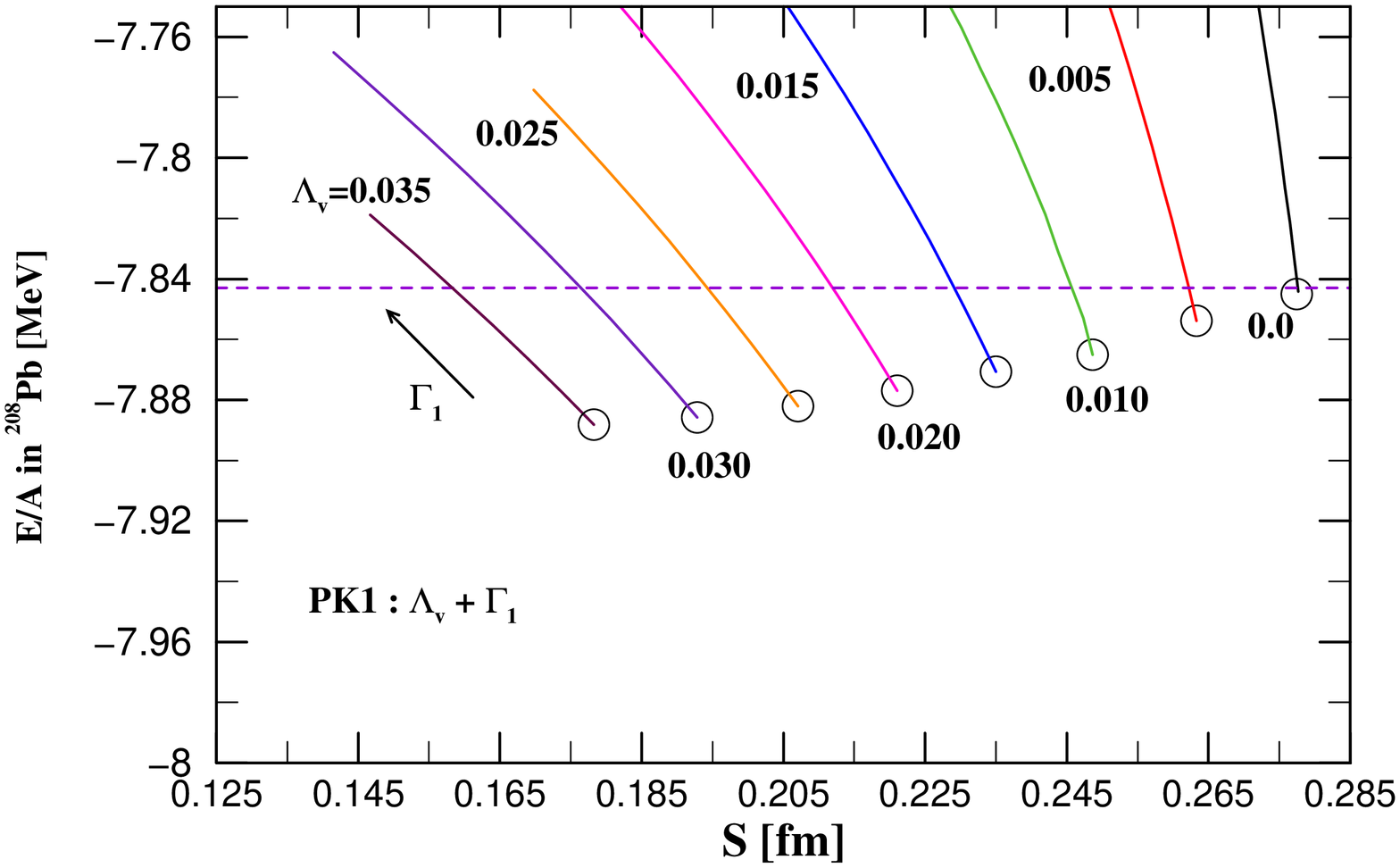}
 \caption{(Color online) Binding energy per nucleon, $E/A$ (in MeV), versus neutron skin
thickness, $S$ (in fm), in $^{208}$Pb for the PK1 effective
interaction for different combinations of $\Lambda_v$ and
$\Gamma_1$, with $\Gamma_2 = 0$. The direction of the arrow
next to the various lines corresponds to increasing values
of $\Gamma_1$. The open circles denote the results for $\Gamma_1 = 0$.
The horizontal dashed line corresponds to the experimental
binding energy per nucleon.}
 \label{fig:pk1-g1+lam}
\end{figure}

We start by considering the PK1 effective interaction and study
the effect of different combinations of $\Lambda_{v}$ and
$\Gamma_{1}$, with $\Gamma_{2} = 0$, as well as different
combinations of $\Lambda_{v}$ and $\Gamma_{1}$, with $\Gamma_{2} =
0$ -- in the total Lagrangian density given by Eq.~(\ref{total})
-- on the neutron skin thickness, neutron and proton rms radii, as
well as neutron and proton density distributions. In order to
understand the role of various coupling constants, we initially
focus on the special case where $\Gamma_{2} = 0$ and consider the
effect of different combinations of $\Lambda_{v}$ and
$\Gamma_{1}$, which are chosen according to the procedure outlined
in Sec.~\ref{sec:extraction}. The value of $g_{\rho}$ is always
chosen so as to reproduce the symmetry energy at an average
density of $k_f$ = 1.15~fm$^{-1}$ corresponding to $a_{asm}$ =
26.08~MeV. In Fig.~\ref{fig:pk1-g1+lam}, the binding energy per
nucleon, $E/A$, versus the neutron skin thickness, $S$, is
presented. Different lines correspond to results with different
$\Gamma_1$ for fixed values of $\Lambda_v$ as indicated in the
figure. The direction of the arrow next to the various lines
corresponds to increasing values of $\Gamma_1$. The open circles
are the results for $\Gamma_1 = 0$. The symbols retain their
meanings in similar figures to follow. The effect of increasing
$\Gamma_1$ is to decrease $S$, while decreasing the absolute value
of the binding energy. With the additional constraint that the
experimental binding energy per nucleon, $(E/A)_{exp}$, is
$-$7.843~MeV (indicated by the dashed line in
Fig.~\ref{fig:pk1-g1+lam}), one can extract values of $S$ which
range from 0.159 to 0.277~fm for different combinations of
$\Gamma_1$ and $\Lambda_v$. More detailed results are listed in
Table \ref{tab:PK11}.

\begin{figure}[htbp]
 \centering
 \includegraphics[height=15cm,angle=-90]{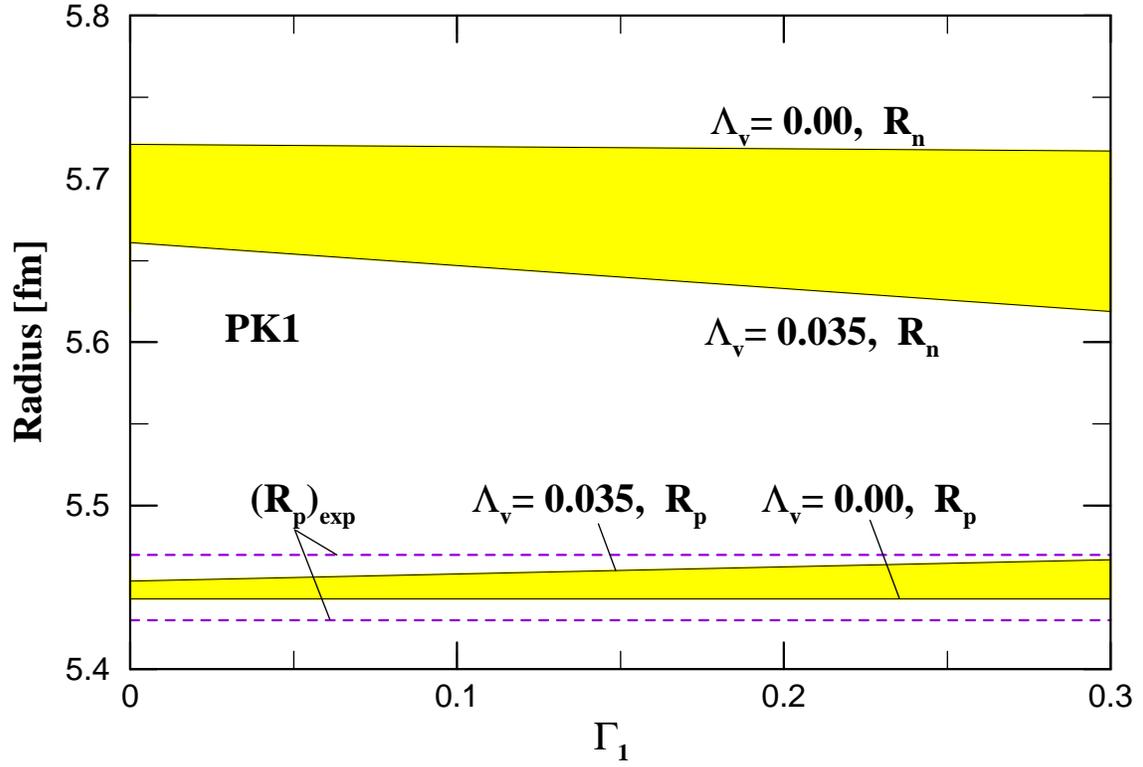}
 \caption{(Color online) Proton and neutron radii, $R_{p}$ and $R_{n}$ respectively,
 in $^{208}$Pb for the PK1 effective interaction for different combinations of
$\Lambda_v$ and $\Gamma_1$, with $\Gamma_2 = 0$. The shaded
regions denote the variation in the values of the radii associated
with different combinations of these parameter sets. Also
indicated, by the dashed horizontal lines, are the lower and upper
limits associated with the experimentally extracted value of the
proton radius.}
 \label{fig:pk1-g1-r}
\end{figure}

\begin{figure}[htbp]
 \centering
 \includegraphics[height=15cm,angle=-90]{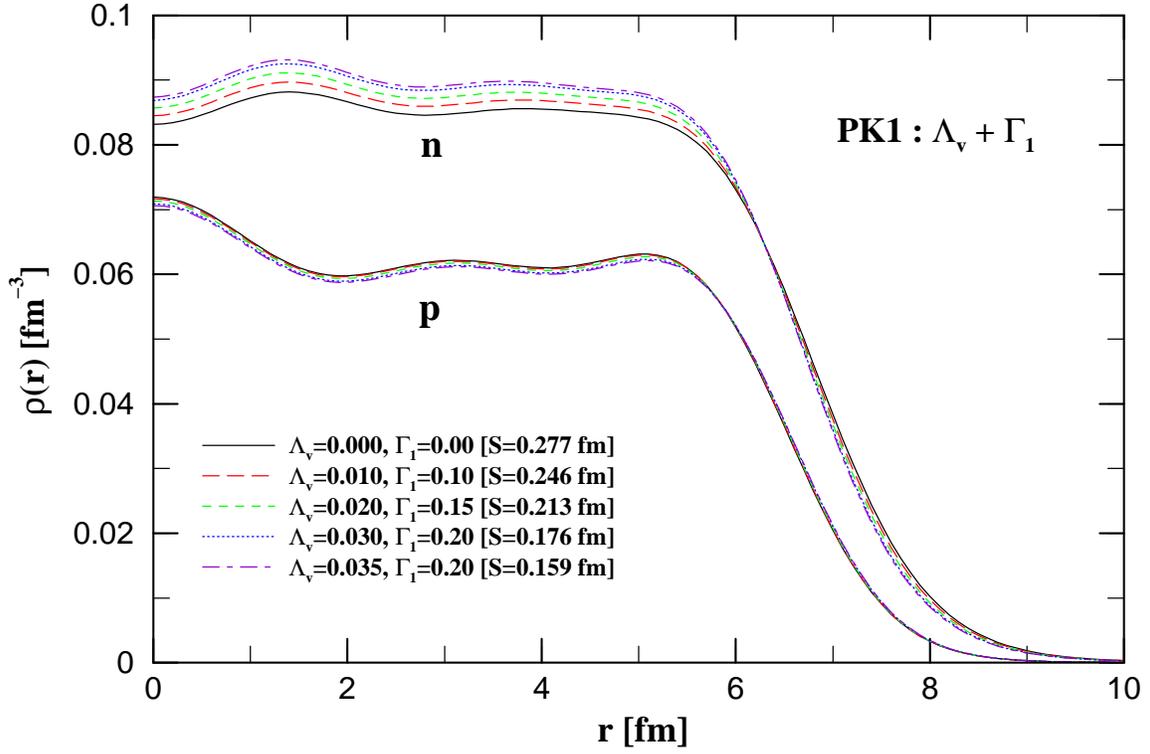}
 \caption{(Color online) Neutron, $n$, and proton, $p$, density distributions (in fm$^{-3}$) in
$^{208}$Pb for the PK1 effective interaction for combinations of
$\Lambda_v$ and $\Gamma_1$, with $\Gamma_2 = 0$, which reproduce
the the experimental binding energy per nucleon for $^{208}$Pb.
The value in square brackets indicates the neutron skin thickness,
$S$, corresponding to a specific paramater set.}
 \label{fig:pk1g1-d-r}
\end{figure}

The accurately measured value of the proton rms radius, $R_p$, of
$^{208}$Pb, namely 5.45 $\pm$ 0.02~fm \cite{Fricke}, places a
stringent constraint on the choice of coupling constants. In
Fig.~\ref{fig:pk1-g1-r}, we also display the neutron and proton
rms radii as functions of different combinations of $\Lambda_v$
and $\Gamma_1$, with $\Gamma_2 = 0$. It is seen that the neutron
rms radius is sensitive to different combinations of $\Gamma_1$
and $\Lambda_v$, whereas the proton rms radius is insensitive and
in good agreement with the experimental value. In
Fig.~\ref{fig:pk1g1-d-r}, the neutron and proton density
distributions have been computed for five different $\Lambda_v$
and $\Gamma_1$ combinations, which reproduce the experimental
binding energy per nucleon for $^{208}$Pb. While combinations of
$\Lambda_v$ and $\Gamma_1$ with small values of $S$ tolerate
larger neutron central densities, the proton density distributions
essentially remain unchanged in all cases.

\begin{figure}[htbp]
 \centering
 \includegraphics[height=15cm,angle=-90]{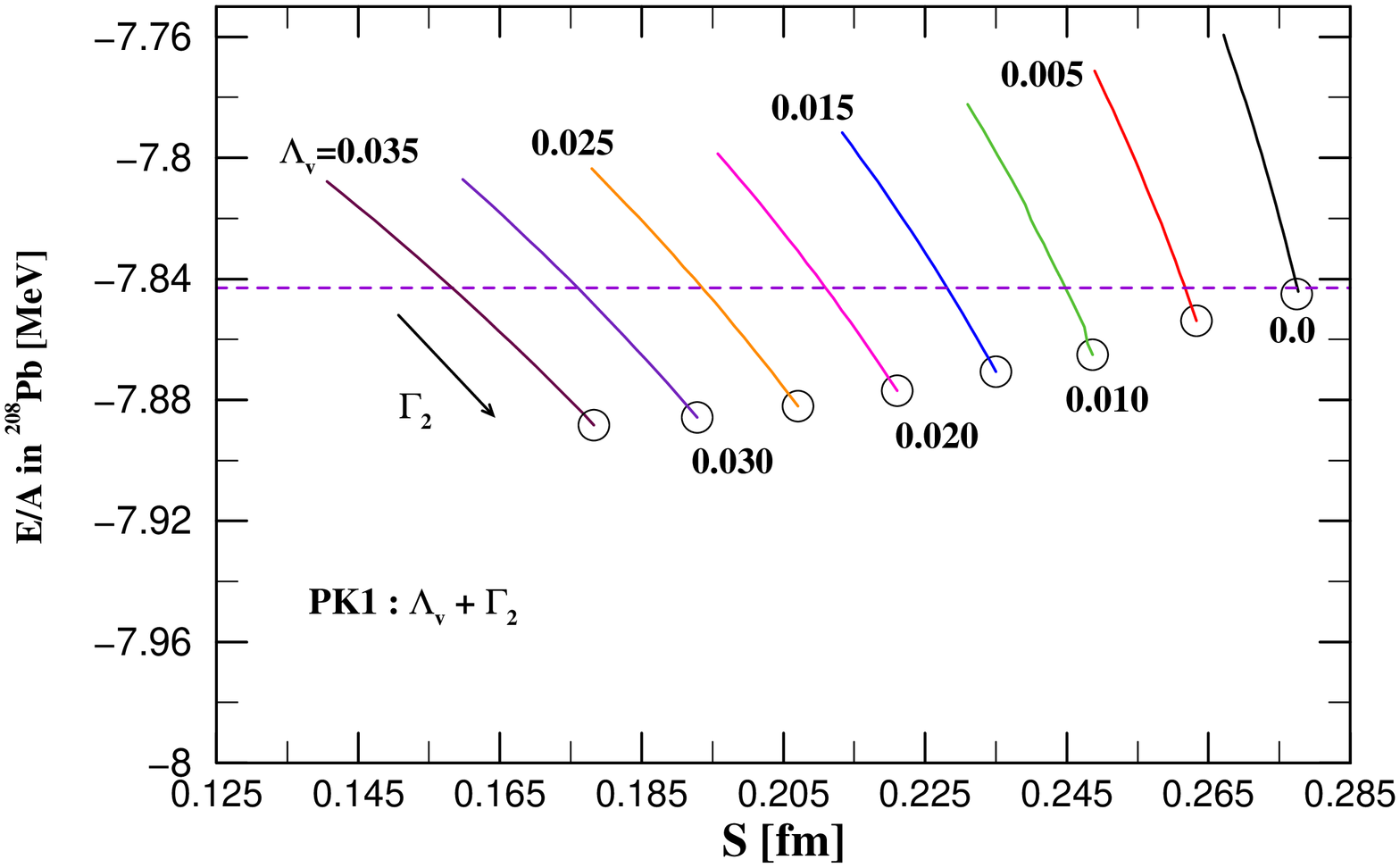}
 \caption{(Color online) Binding energy per nucleon, $E/A$ (in MeV), versus the neutron skin
thickness, $S$ (in fm), in $^{208}$Pb for the PK1 effective
interaction for different combinations of $\Lambda_v$ and
$\Gamma_2$, with $\Gamma_1 = 0$. The direction of the arrow
next to the various lines corresponds to increasing values
of $\Gamma_2$. The open circles denote the results for $\Gamma_2 = 0$.
The horizontal dashed line corresponds to the experimental
binding energy per nucleon.
}
 \label{fig:pk1-g2+lam}
\end{figure}

Next, we consider the PK1 effective interaction and study the
effect of different combinations of $\Lambda_{v}$ and
$\Gamma_{2}$, for $\Gamma_{1} = 0$, on $S$. The binding energy per
nucleon, $E/A$, versus $S$ is shown in Fig.~\ref{fig:pk1-g2+lam}.
The symbols retain the same meanings as in
Fig.~\ref{fig:pk1-g1+lam}, except that the roles of $\Gamma_2$ and
$\Gamma_1$ are exchanged. Constraining the values of $S$ via the
experimental binding energy per nucleon, yields essentially the
same range of values for $S$ as was obtained in
Fig.~\ref{fig:pk1-g1+lam}. The main difference between
Figs.~\ref{fig:pk1-g1+lam} and \ref{fig:pk1-g2+lam} lies in the
fact that the values of $\Gamma_2$ are negative and the absolute
values thereof are larger than $\Gamma_1$ for fixed values of
$\Lambda_v$, and for the same values of the neutron skin
thickness. More detailed results for different $\Gamma_2$ and
$\Lambda_v$ combinations are listed in Table \ref{tab:PK12}. Once
again, we mention the fact that the proton rms radius is well
constrained by the experimental data.

Previous RMF models predicted values of $S$ in $^{208}$Pb which
ranged from 0.2 to 0.3~fm \cite{Typel,Furn}. In our present
analysis, we obtain values of $S$ ranging from 0.159 to 0.277~fm,
for different $\Lambda_v$ and $\Gamma_1$ (or $\Gamma_2$)
combinations for the PK1 effective interaction, with the binding
energy and proton rms radius both constrained by data. More
specifically, our new models yield a minimum value of $S$ =
0.159~fm for $\Lambda_v$ = 0.035 and $\Gamma_1$ = 0.2 (or
$\Gamma_2$ = $-$0.55) in comparison with $S$ = 0.16 $\pm$ 0.02~fm
given by modern Skryme Hartree-Fock models \cite{Brown}.

\begin{figure}[htbp]
 \centering
 \includegraphics[height=8cm,angle=-90]{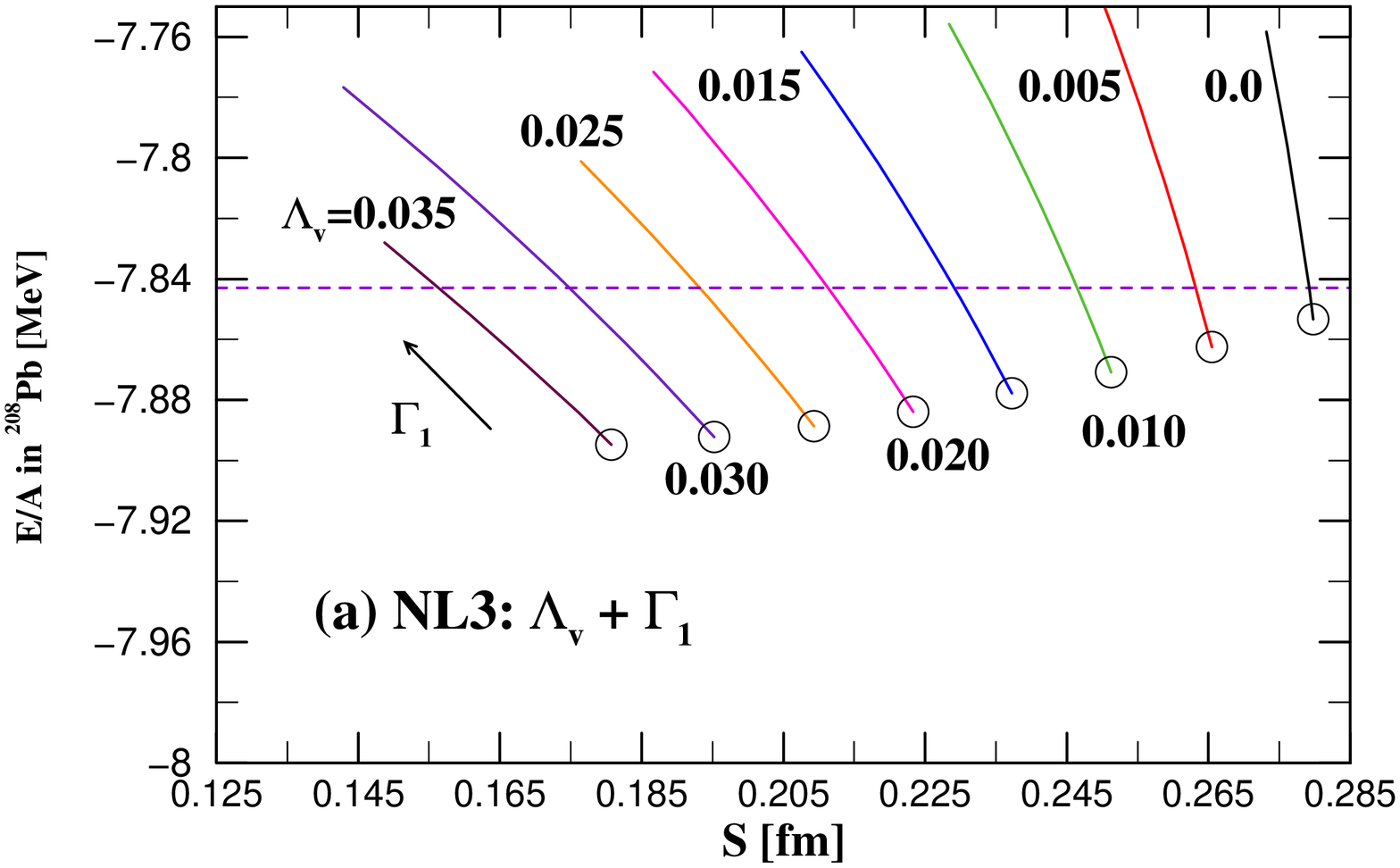}
 \includegraphics[height=8cm,angle=-90]{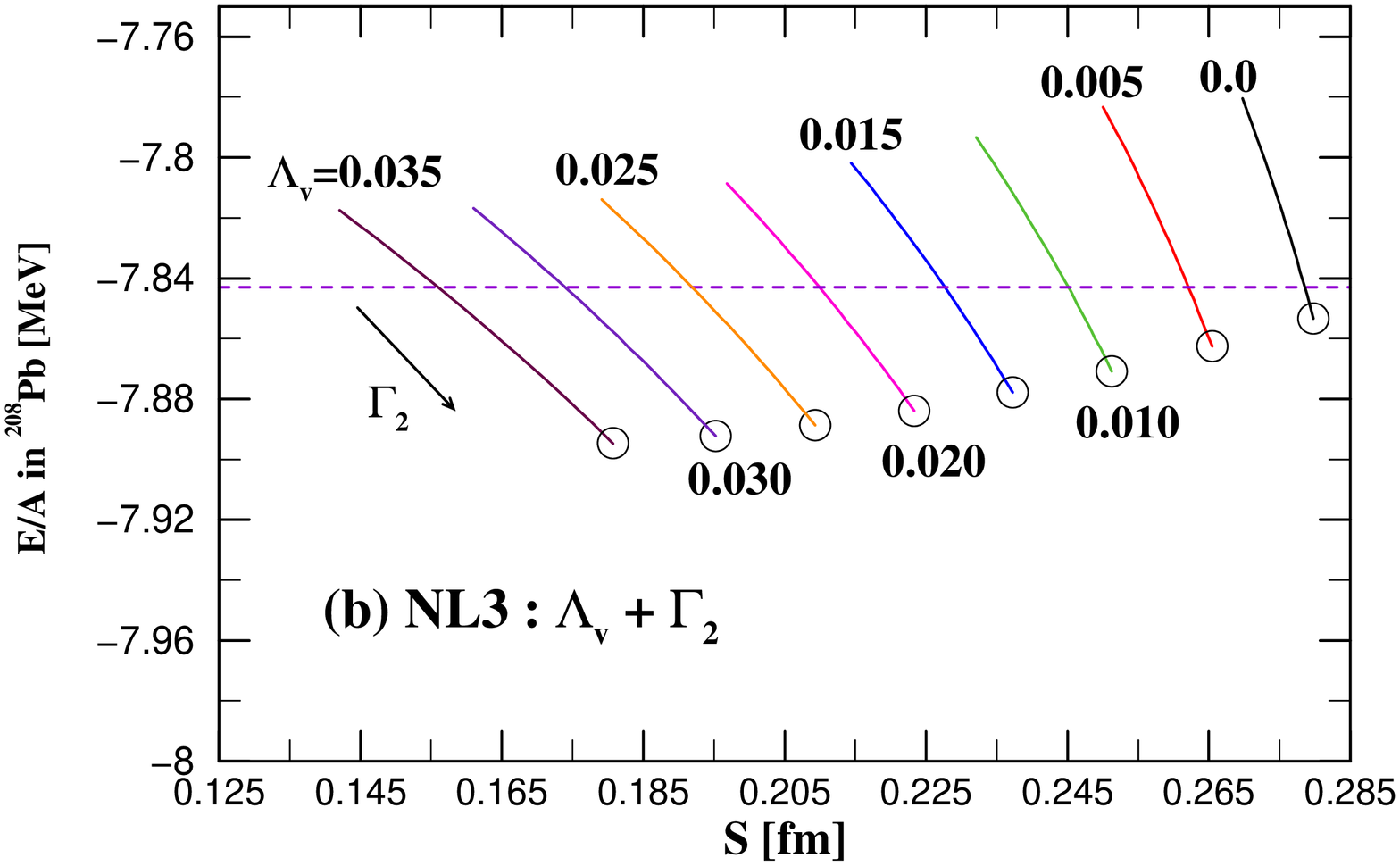}
 \caption{(Color online) Binding energy per nucleon, $E/A$ (in MeV),
versus neutron skin thickness, $S$ (in fm), in $^{208}$Pb for the
NL3 effective interaction for different combinations of (a)
$\Lambda_v$ and $\Gamma_1$ (left panel), with $\Lambda_{2} = 0$,
and (b) $\Lambda_v$ and $\Gamma_2$ (right panel), with
$\Lambda_{1} = 0$. The direction of the arrows next to the various
lines corresponds to increasing values of $\Gamma_1$
($\Gamma_{2}$). The open circles denote the results for $\Gamma_1
= 0$ ($\Gamma_{2} = 0$). The horizontal dashed line corresponds to
the experimental binding energy per nucleon.} \label{fig:PN}
\end{figure}

We now repeat the above analysis for the NL3 effective
interaction. In Fig.~\ref{fig:PN}, the binding energy per nucleon,
$E/A$, versus $S$ is shown for different combinations of (a)
$\Lambda_v$ and $\Gamma_1$ (left panel), with $\Gamma_{2} = 0$,
and (b) for different combinations of $\Lambda_v$ and $\Gamma_2$
(right panel), with $\Gamma_{1} = 0$. Constraining the values of
$S$ via the experimental binding energy per nucleon, yields a
range of values for $S$ similar to those associated with the PK1
effective interaction. The minimum value of $S$ is 0.155~fm and
this corresponds to the $\Lambda_v$ = 0.035 and $\Gamma_1$ = 0.25,
with $\Gamma_2$ = 0, combination. More detailed data regarding the
parameter sets and associated observables are listed in Tables
\ref{tab:NL31} and \ref{tab:NL32}.

\begin{figure}[htbp]
 \centering
 \includegraphics[height=8cm,angle=-90]{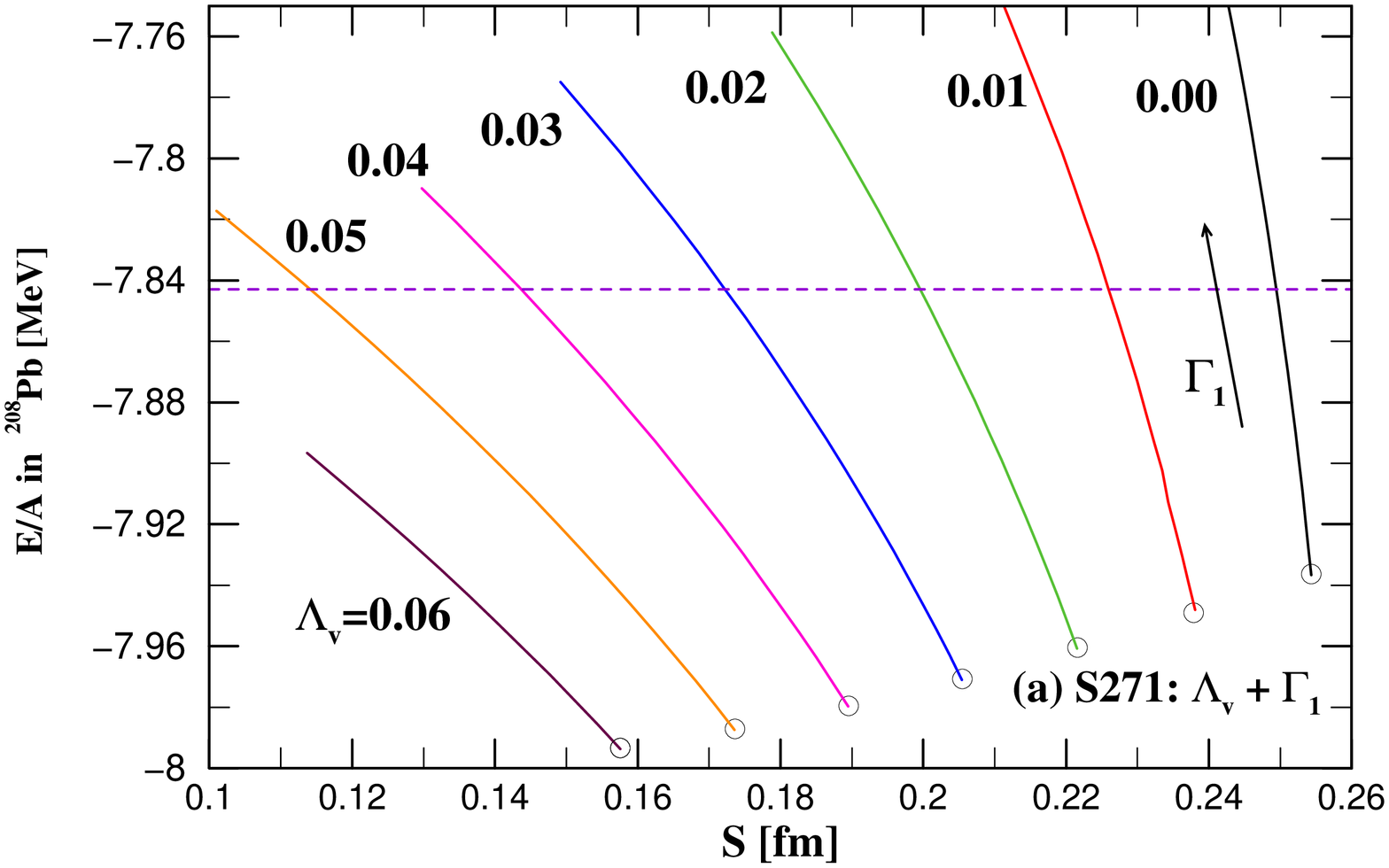}
 \includegraphics[height=8cm,angle=-90]{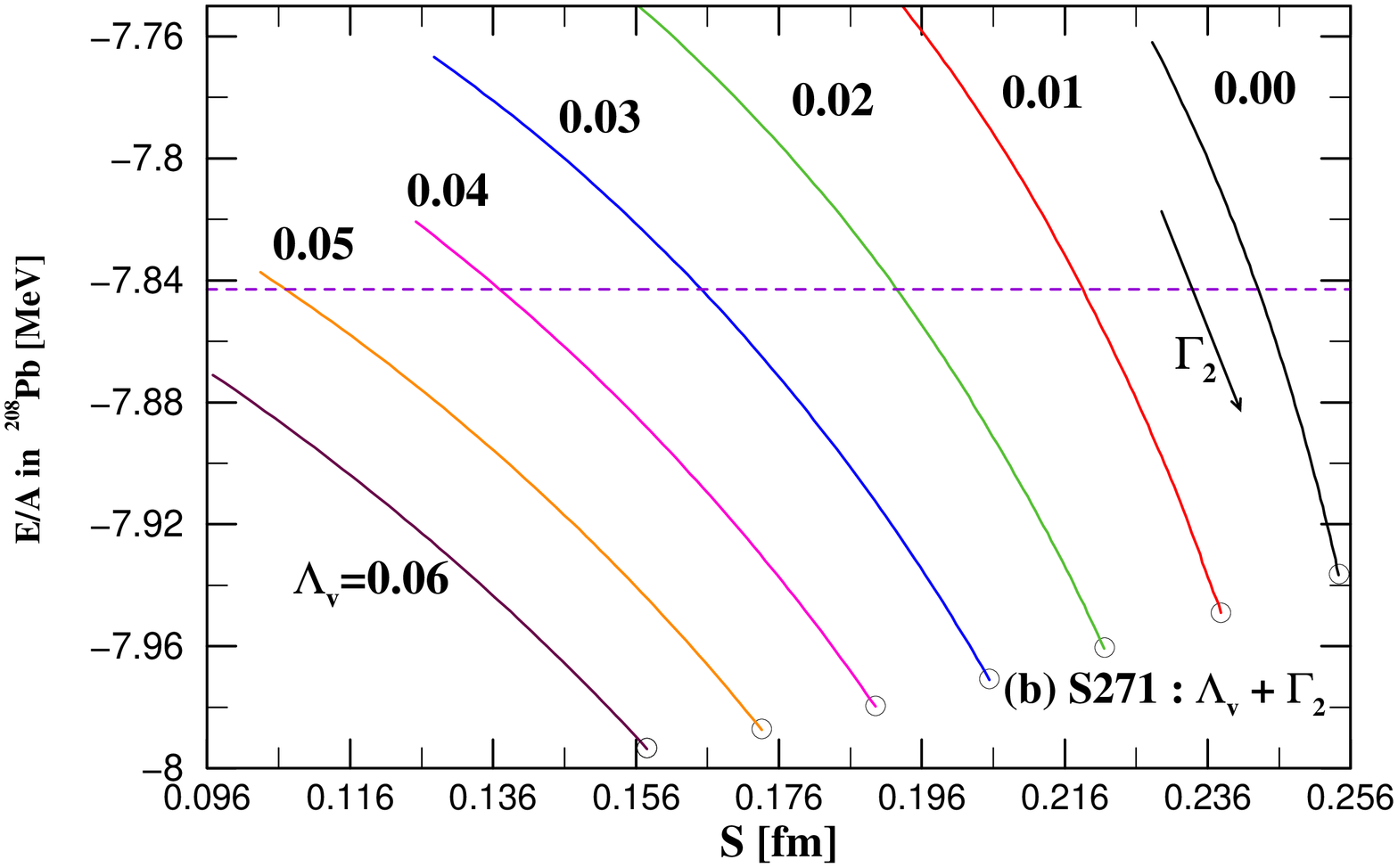}
 \includegraphics[height=8cm,angle=-90]{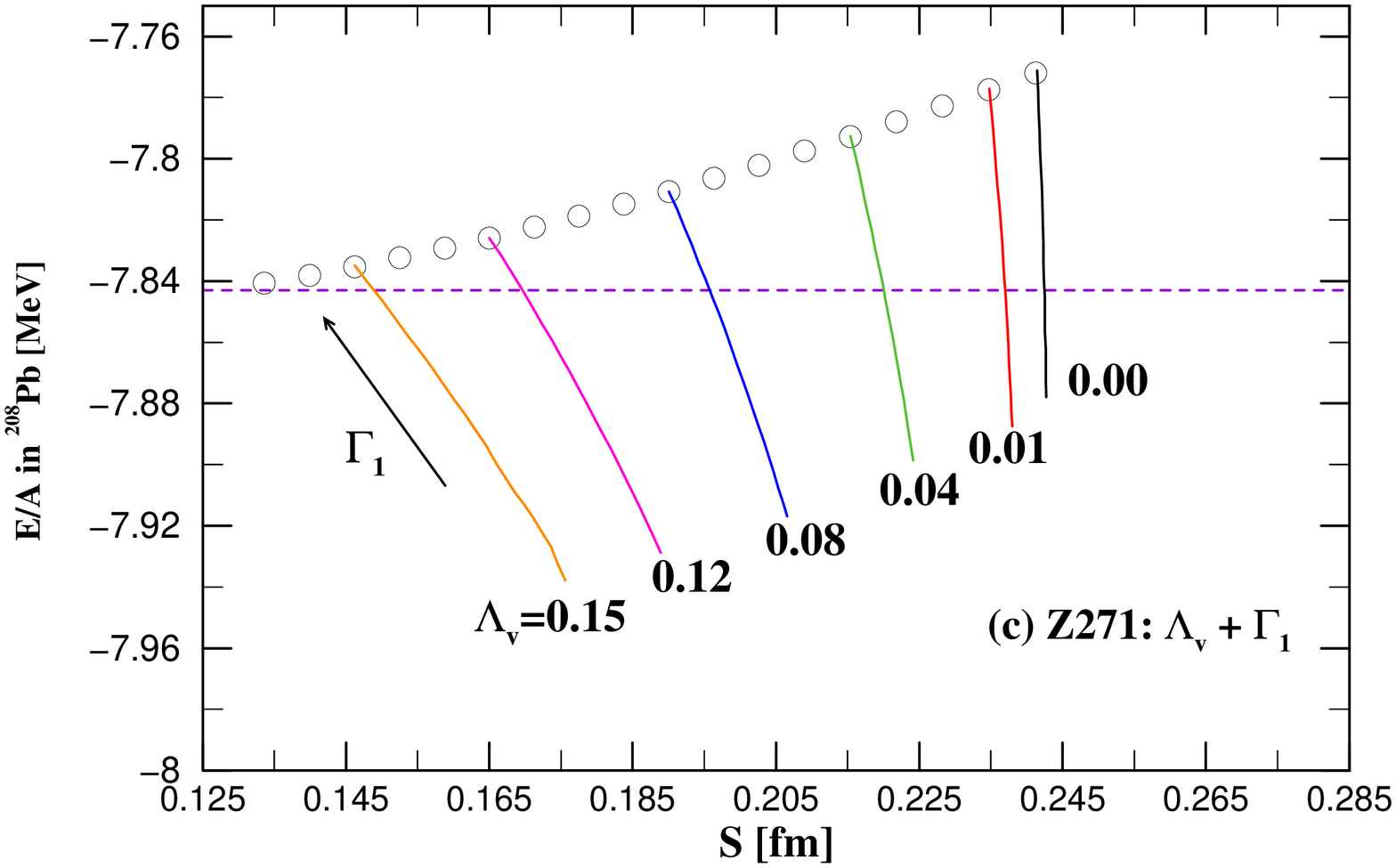}
 \includegraphics[height=8cm,angle=-90]{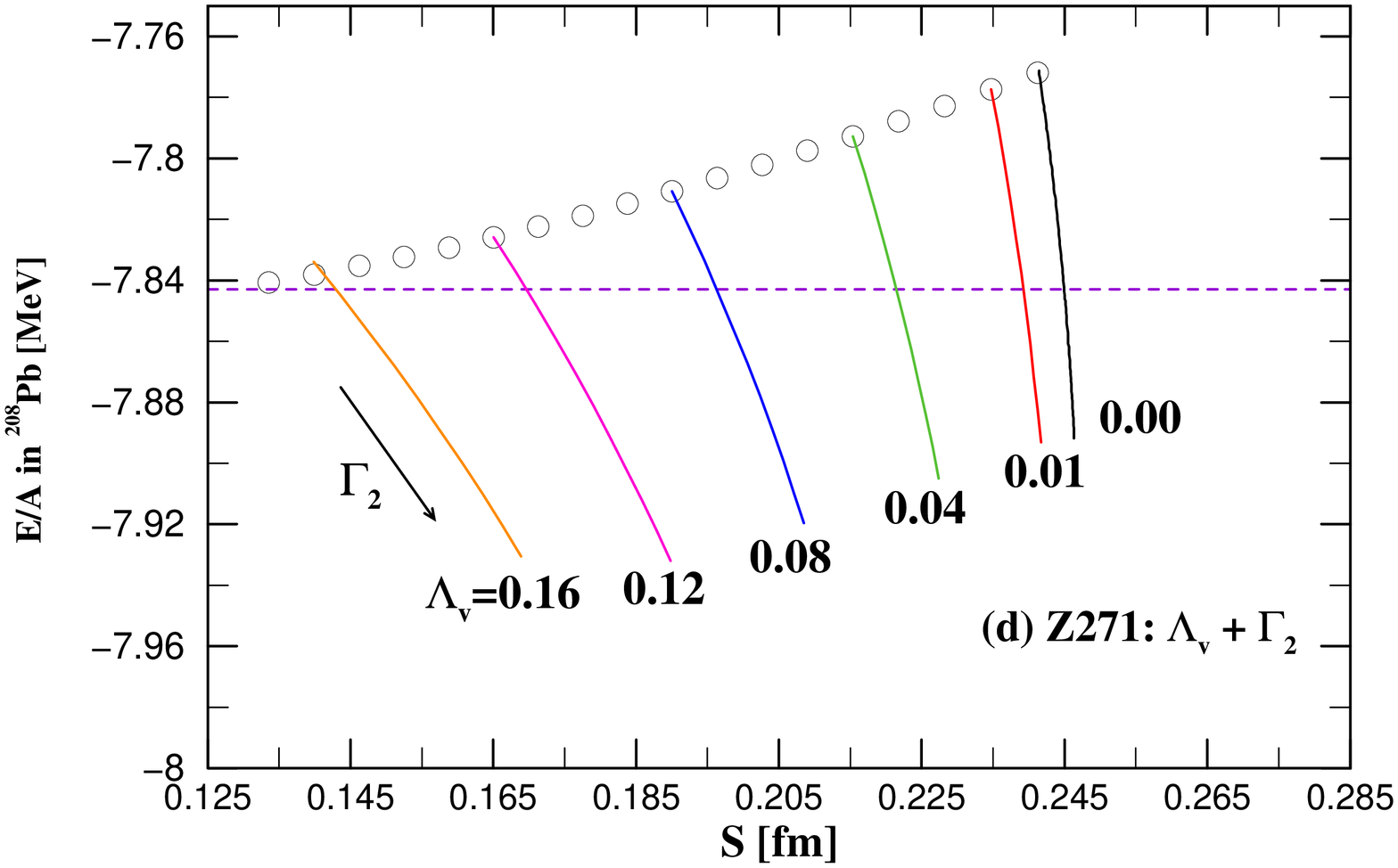}
 \caption{(Color online) Binding energy per nucleon, $E/A$ (in MeV),
versus neutron skin thickness, $S$ (in fm), in $^{208}$Pb for the
(a) S271 effective interactions for different combinations of
$\Lambda_v$ and $\Gamma_1$ (left panel), with $\Lambda_{2} = 0$;
(b) S271, $\Lambda_v$ and $\Gamma_2$ (right panel), with
$\Lambda_{1} = 0$; (c) Z271, $\Lambda_v$ and $\Gamma_1$ (left
panel), with $\Lambda_{2} = 0$; (d) Z271, $\Lambda_v$ and
$\Gamma_2$ (right panel), with $\Lambda_{1} = 0$. The direction of
the arrows next to the various lines corresponds to increasing
values of $\Gamma_1$ ($\Gamma_{2}$). The open circles denote the
results for $\Gamma_1 = 0$ ($\Gamma_{2} = 0$). The horizontal
dashed line corresponds to the experimental binding energy per
nucleon.}
 \label{fig:SZ271}
\end{figure}

Next, we study the effect of adding our new terms on two
additional effective interactions, namely the S271 and Z271
parameterizations of RMF \cite{H1,H3}. The original parameters for
these interactions were extracted by fitting to the properties of
nuclear matter and the experimentally determined proton radius of
$^{208}$Pb. The addition of the $\Lambda_{v}$ term, given by
Eq.~(\ref{LHP}), to the Lagrangian Eq.~(\ref{lagrangian}), reduces
the values of the neutron skin thickness in $^{208}$Pb for both
the S271 and Z271 effective interactions, while sacrificing good
fits to the experimental binding energy per nucleon. The latter
was the result of Ref.~\cite{H3}, the points of which are
indicated by open circles in Fig.~\ref{fig:SZ271}. The effect of
different combinations [in the Lagrangian given by
Eq.~(\ref{total})] of $\Lambda_{v}$ and $\Gamma_{1}$, with
$\Gamma_{2} = 0$,  and various combinations of $\Lambda_{v}$ and
$\Gamma_{2}$, with $\Gamma_{1} = 0$, on $S$ is displayed in the
different panels in Fig.~\ref{fig:SZ271}. Constraining the values
of $S$ via the experimental binding energy per nucleon for the
S271 effective interaction (see upper panels in
Fig.~\ref{fig:SZ271}), yields values of $S$ which range from 0.106
to 0.258~fm, for different combinations of $\Gamma_1$ ($\Gamma_2$)
and $\Lambda_v$. A slight narrower range is obtained for the Z271
effective interaction, namely $S$ = 0.134 $\sim$ 0.241~fm: see
lower panels in Fig.~\ref{fig:PN}. The main difference between the
S271 and Z271 effective interactions is that the signs of
$\Gamma_1$ and $\Gamma_2$ are opposite at points corresponding to
the experimental binding energy per nucleon. More detailed data on
$S$ versus the binding energy per nucleon for different $\Gamma_1$
and $\Lambda_v$ combinations for S271 and Z271 are presented in
Tables \ref{tab:S2711} and \ref{tab:Z2711}. As is the case with
the PK1 and NL3 effective interactions, we have checked that the
proton rms radius of $^{208}$Pb agrees well with the experimental
data.

\begin{figure}[htbp]
 \centering
 \includegraphics[height=15cm,angle=-90]{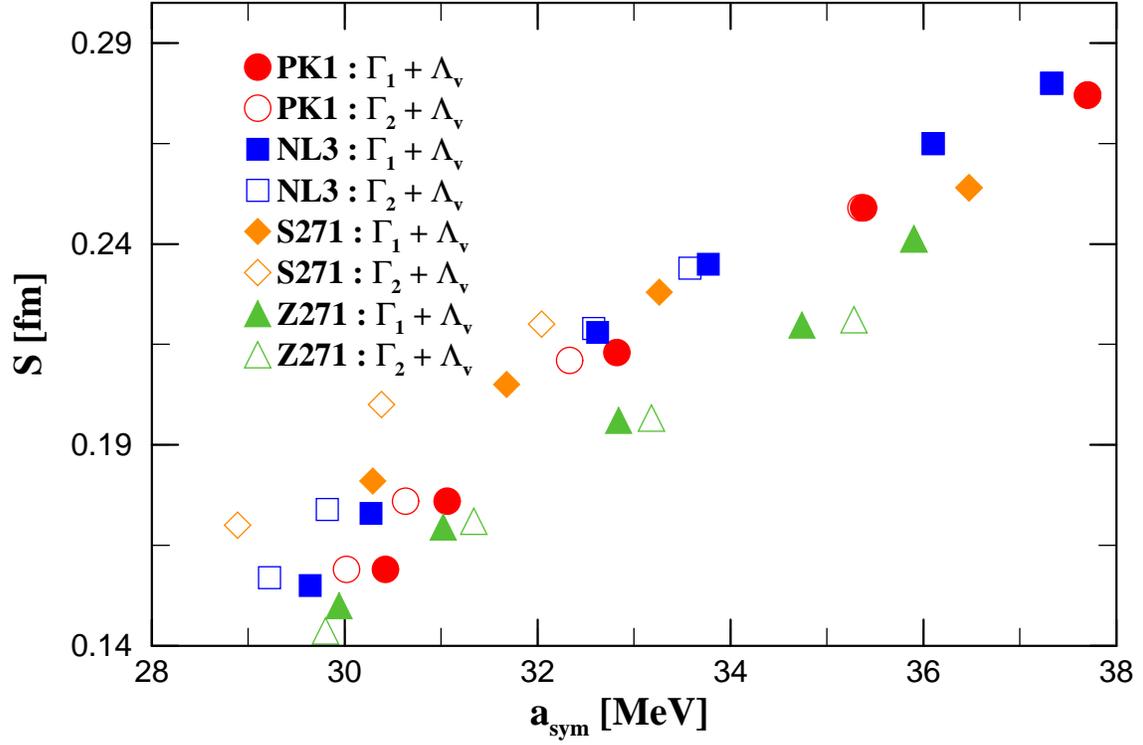}

 \caption{(Color online) Neutron skin thickness, $S$ (in fm), in $^{208}$Pb versus
the symmetry energy, $a_{sym}$ (in MeV), of nuclear matter at
saturation density for the PK1, NL3, S271, and Z271 effective
interactions for various combinations of $\Lambda_v$ and
$\Gamma_1$ ($\Gamma_2$), which reproduce the experimental binding
energy per nucleon for $^{208}$Pb.}
 \label{fig:S-asym}
\end{figure}

Previously it has been demonstrated that, for the original RMF
models of interest to this paper, a linear relationship exists
between the neutron skin thickness for $^{208}$Pb and symmetry
energy of nuclear matter at saturation density
\cite{Furn,Brown,Diep}. After the inclusion of our new terms,
given by Eqs.~(\ref{L1}) and (\ref{L2}), we study this relation
for each of PK1, NL3, S271 and Z271 effective interactions: the
results are presented in Fig.~\ref{fig:S-asym}. For this study we
only take combinations of $\Lambda_{v}$ and $\Gamma_{1}$
($\Gamma_{2}$) which reproduce both the experimental binding
energy as well as the experimental proton rms radius of
$^{208}$Pb. Indeed, we also observe a linear relation between the
neutron skin thickness, $S$ (in fm), and the nuclear symmetry
energy, $a_{sym}$ (in MeV), which can be expressed as:
\beqn\label{fit} S \approx -0.283\ +\ 0.015\ a_{sym}\,. \eeqn At
the saturation point, the symmetry energy, $a_{sym}$, for the
various interactions, ranges from 29 to 38~MeV and the neutron
skin thickness, $S$, ranges from 0.14 to 0.28~fm.

The addition of our new higher order correction terms, expressed
via Eqs.~(\ref{L1}) and (\ref{L2}), to existing RMF models can be
associated with some kind of multi-meson exchange processes
occurring in the inner (higher density) region of nuclei. In
low-energy nucleon-nucleon scattering models, the exchange of
mesons with an effective mass of less than 1~GeV seems to be the
dominant contribution to the scattering amplitudes. After
one-meson exchange, the inclusion of two-meson exchange at low
energies has been demonstrated to provide a major improvement in
the theoretical description of nucleon-nucleon scattering data
\cite{Th}. In the inner region of a heavy nucleus such as
$^{208}$Pb, one could imagine that multi-meson exchange processes,
expressed via Eqs.~(\ref{L1}) and (\ref{L2}), could play an
important role in modifying the short range behaviour of the
nucleon-nucleon interaction, by softening the density dependence
of the symmetry energy at high density (the inner part of nuclei)
and thus leading to smaller values of neutron skin thickness $S$
compared to previous interactions which excluded our new
correction terms.

In principle one could include the isospin-dependent nuclear force
associated with the long-range pseudoscalar pion. However, because
of parity conservation, the pion cannot contribute in a mean-field
treatment of the ground state of spherical nuclei as $^{208}$Pb.
Moreover, what determines the neutron skin thickness seems to be
the neutron distribution in the inner core of nuclei, which in
turn is more sensitive to the short-range part of the nuclear
force. The above arguments suggest that pion contributions do not
play an important role in determining the neutron skin thickness
in $^{208}$Pb.

\begin{figure}[htbp]
 \centering
 \includegraphics[height=15cm,angle=-90]{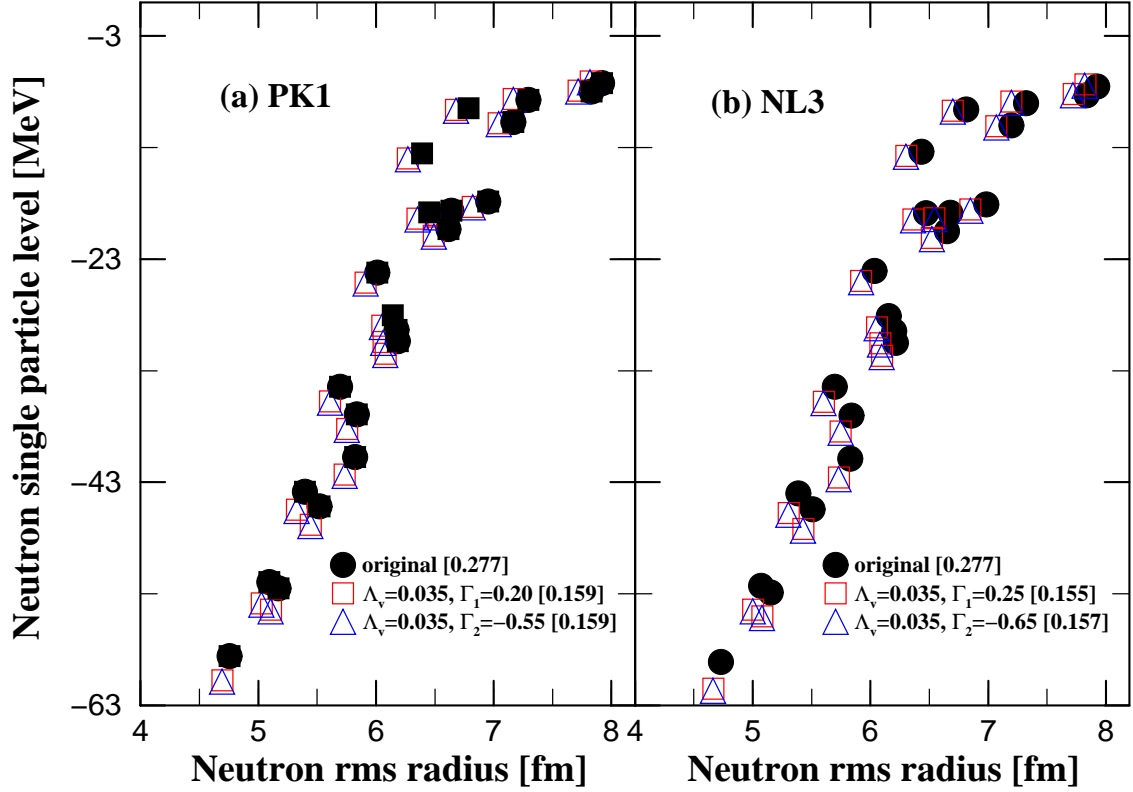}
 \caption{(Color online) Neutron single particle energy levels (in
MeV) versus their corresponding neutron rms radii (in fm) in
$^{208}$Pb for the (a) PK1 and (b) NL3 effective interactions:
original one ($\Lambda_{v} = \Gamma_{1} = \Gamma_{2} = 0$) denoted
by filled circles, as well as for specific combinations of
$\Lambda_{v}$ and $\Gamma_{1}$ (open squares), with $\Gamma_{2} =
0$, and also $\Lambda_{v}$ and $\Gamma_{2}$ (open triangles), with
$\Gamma_{1} = 0$ which yield a minimum value for $S$ (in square
brackets) for the corresponding combinations of parameters.}
 \label{fig:pk1-Nl3-level}
\end{figure}

Since the parity-violating electron scattering experiment at
Jefferson Lab can only provide information about the neutron
radius, and hence also $S$, in $^{208}$Pb, it is also important to
study the consequences of an exact measurement of $S$ on other
properties of this nucleus. In particular, we now observe the
influence of different $\Gamma_1$ ($\Gamma_2$) and $\Lambda_v$
combinations (which reproduce both the experimental binding energy
as well as the experimental proton rms radius of $^{208}$Pb), for
both PK1 and NL3 effective interactions, to neutron single
particle energy levels versus their corresponding neutron rms
radii in $^{208}$Pb: the results are displayed in
Fig.~\ref{fig:pk1-Nl3-level}. We only consider the PK1 and NL3
effective interactions since they have been used to provide an
excellent description of nuclei near and far from the valley of
$\beta$ stability. First we consider the original PK1 effective
interaction ($\Lambda_{v} = \Gamma_{1} = \Gamma_{2} = 0$) which
yields $S$ = 0.277~fm. The corresponding neutron single particle
energy levels are denoted by filled circles in the left panel of
Fig.~\ref{fig:pk1-Nl3-level}. We now compare these single particle
energy levels to the values corresponding to those combinations of
$\Lambda_v$ and $\Gamma_1$, with $\Gamma_2 = 0$, and $\Lambda_v$
and $\Gamma_2$, with $\Gamma_1 = 0$, which yield a minimum value
for $S$. In particular, the combination of $\Lambda_v$ = 0.035 and
$\Gamma_1$ = 0.2, with $\Gamma_2 = 0$, yields a mininum value of
$S = 0.159$~fm for the PK1 effective interaction and the
corresponding single particle energy levels are denoted by open
squares in Fig.~\ref{fig:pk1-Nl3-level}. Also the combination of
$\Lambda_v$ = 0.035 and $\Gamma_2$ = $-$0.55, with $\Gamma_{1} =
0$, yields a mininum value of $S = 0.159$~fm: the corresponding
single particle energy levels are denoted by open triangles in
Fig.~\ref{fig:pk1-Nl3-level}. The quantum numbers associated with
each of the levels are the same as those associated with usual RMF
predictions: see for example Ref.~\cite{Long}. The effect of
decreasing $S$ is to deepen each single particle energy level and
to decrease the corresponding neutron rms radius. More
specifically, we see the addition of our new higher order
correction terms is to deepen the deep-lying states more than the
low-lying states. This, in turn, suggests the possibility that the
effect of these new terms is to modify the short range behavior of
the nucleon-nucleon interaction. In the right panel of
Fig.~\ref{fig:pk1-Nl3-level}, we observe the effect of adding our
new terms on the neutron single particle energy levels and their
corresponding neutron rms radii for the NL3 effective interaction:
the results and discussion are similar to those associated with
the PK1 effective interaction.

\subsection{Correlation between neutron skin thickness and the radius of a neutron star}

We now discuss the effect of our new terms, given by
Eqs.~(\ref{L1}) and (\ref{L2}), on the radius, $R$, of a 1.4
solar-mass neutron star and also study its correlation  with the
neutron skin thickness, $S$, in $^{208}$Pb for the PK1, NL3, S271
and Z271 effective interactions. In partciular, we consider the
following combinations of parameters: (1) $\Lambda_v$, $\Gamma_1$
and $\Gamma_2$ terms separately, (2) $\Lambda_v$ and $\Gamma_1$
terms, with $\Gamma_2$ = 0, and(3) $\Lambda_v$ and $\Gamma_2$
terms, with $\Gamma_1$ = 0.

\begin{figure}[htbp]
\centering
\includegraphics[height=15cm,angle=-90]{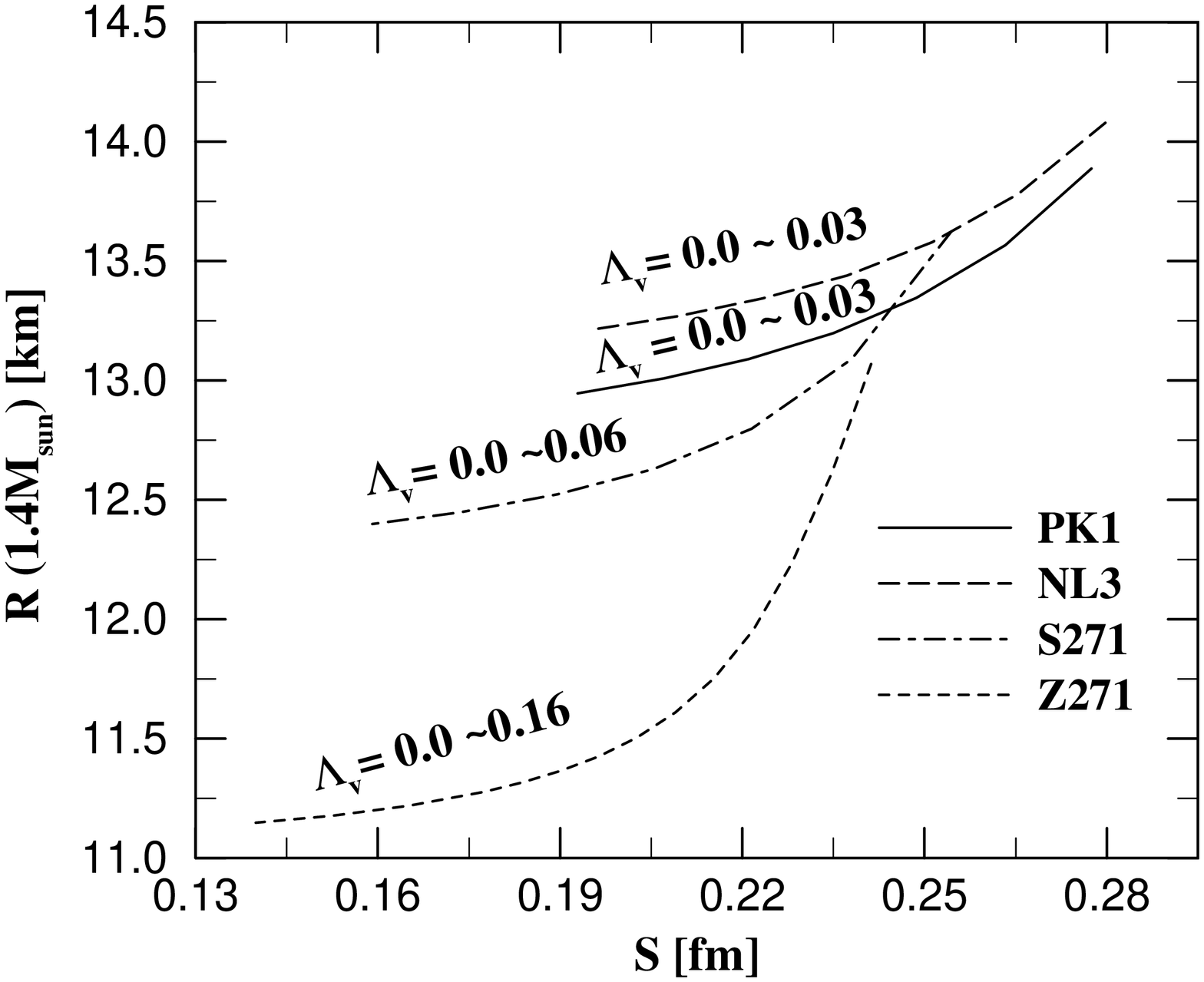}
\caption{The radius, $R$ (in km), of a 1.4 solar-mass neutron star
versus the neutron skin thickness, $S$ (in fm), in $^{208}$Pb for
the PK1, NL3, S271, and Z271 effective interactions for different
ranges of values of $\Lambda_v$, indicated next to the various line types.} \label{fig:lambdav_all.eps}
\end{figure}

After adding the $\Lambda_v$ term, given by Eq.~(\ref{LHP}), to
the Lagrangian given by Eq.~(\ref{lagrangian}), the relationship
between $R$ and $S$ is displayed in
Fig.~\ref{fig:lambdav_all.eps}: different line types correspond to
different interactions. For all the interactions, values of $R$
increase with increasing values of $S$. However, different
interactions give different values of $R$ for the same value of
$S$. For example, for $S$ = 0.21~fm, values of $R$ vary from
13.27~km for the NL3 effective interaction to 11.63~km for the
Z271 effective interaction. Therefore, the radius of a 1.4
solar-mass neutron star is not uniquely constrained by a
measurement of the neutron skin thickness, as has already been
pointed out by Horowitz and Piekarewicz \cite{H3}. The reason for
this is that $S$ depends on the equation of state (EOS) near or
below saturation density where most RMF models are calibrated to
successfully describe the properties of nuclear matter and finite
nuclei, whereas the neutron star radius is mainly sensitive to the
EOS for higher densities. Hence, different RMF models give
different predictions for the properties of neutron stars, and as
such one cannot expect to reliably extrapolate any of these models
to extreme conditions of isospin and density. However, combining
separate measurements of $S$ and $R$ can provide valuable
information on the EOS at low and high densities. For example,
relatively large values of $S$ suggest a stiff low-density EOS,
while relatively small values of $R$ imply a soft high density
EOS, and vice versa.

\begin{figure}[htbp]
\centering
\includegraphics[height=15cm,angle=-90]{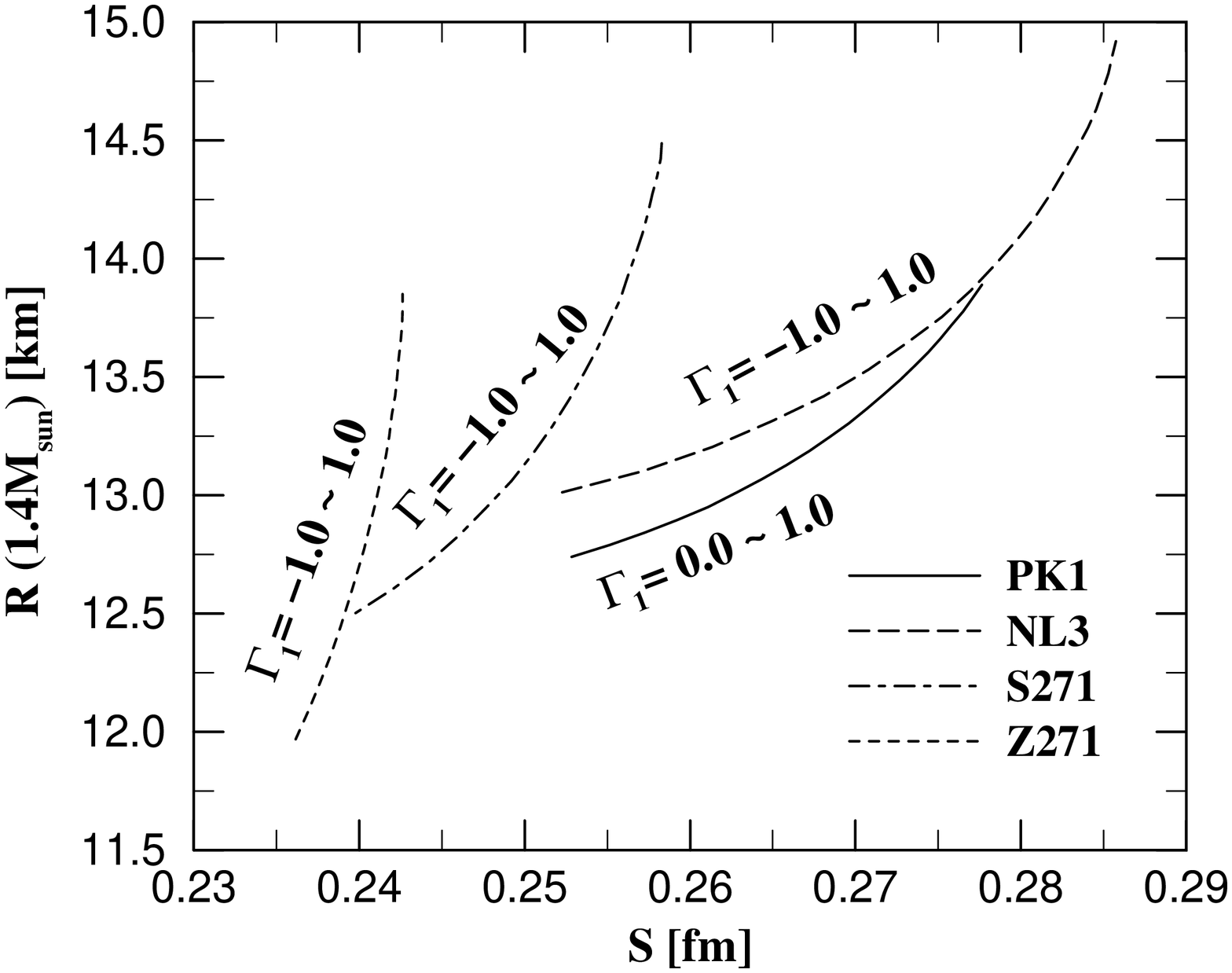}
\caption{The radius, $R$ (in km), of a 1.4 solar-mass neutron star
versus the neutron skin thickness, $S$ (in fm), in $^{208}$Pb for
the PK1, NL3, S271, and Z271 effective interactions
for different ranges of values of $\Gamma_1$, indicated
next to the various line types.} \label{fig:gamma1_all.eps}
\end{figure}

In Fig.~\ref{fig:gamma1_all.eps}, we observe the effect of the
$\Gamma_{1}$ term, given by Eq.~(\ref{L1}), with $\Lambda_{v} =
\Gamma_{2} = 0$, on the correlation between $S$ and $R$. As
expected, $R$ increases with an increasing $S$, and different
models give different relationships between $R$ and $S$. The
results for the PK1 and NL3 effective interactions are generally
similar, but the value of $R$ given by PK1 is smaller than that of
NL3 for the same value of $S$. The S271 and Z271 effective
interactions produce smaller neutron skin thicknesses, and also
$R$ increases more rapidly with $S$ compared to PK1 and NL3. For
example, for $\Gamma_1 = 0$, $S \simeq$ 0.280, 0.278, 0.254,
0.241~fm, and the corresponding values of $R$ are 14.08, 13.89,
13.62, 13.07~km for the NL3, PK1, S271 and Z271 effective
interactions, respectively. As the value of $\Gamma_1$ is
increased to 1.0, the neutron skin thickness decreases to 0.252,
0.253, 0.240, 0.236~fm, and the corresponding values of $R$
decrease to 13.01, 12.74, 12.50, 11.97~km for the NL3, PK1, S271
and Z271 effective interactions, respectively.

\begin{figure}[htbp]
\centering
\includegraphics[height=15cm,angle=-90]{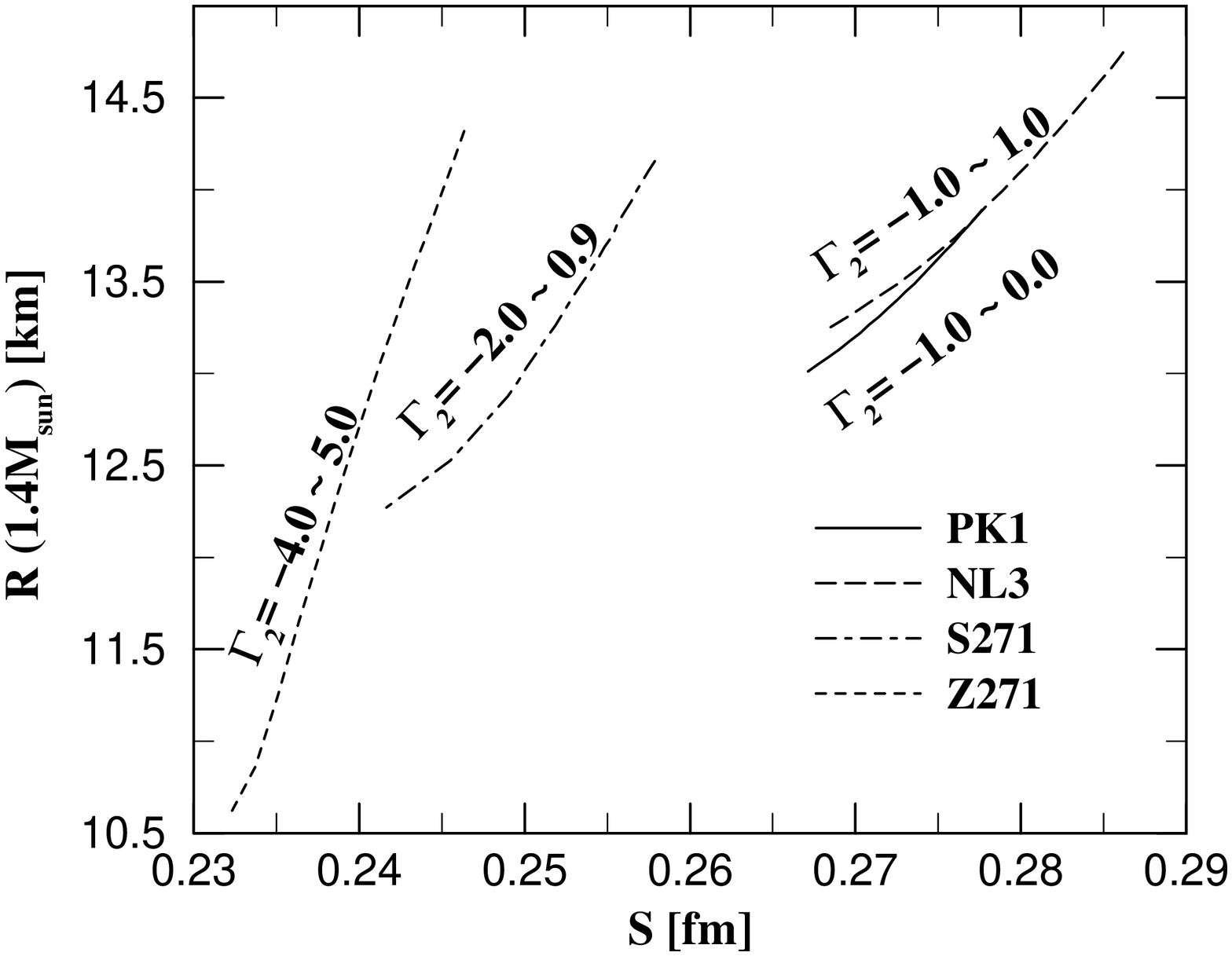}
\caption{The radius, $R$ (in km), of a 1.4 solar-mass neutron star
versus the neutron skin thickness, $S$ (in fm), in $^{208}$Pb for
the PK1, NL3, S271, and Z271 effective interactions
for different ranges of values of $\Gamma_2$, indicated
next to the various line types.} \label{fig:gamma2_all.eps}
\end{figure}

In Fig.~\ref{fig:gamma2_all.eps}, we observe the effect of the
$\Gamma_{2}$ term, given by Eq.~(\ref{L2}), with
$\Lambda_{v} = \Gamma_{1} = 0$, on the correlation between $S$ and $R$: the
results and discussion are very similar to those associated
with Fig.~\ref{fig:gamma1_all.eps}.

\begin{figure}[htbp]
\centering
\includegraphics[width=6cm,angle=-90]{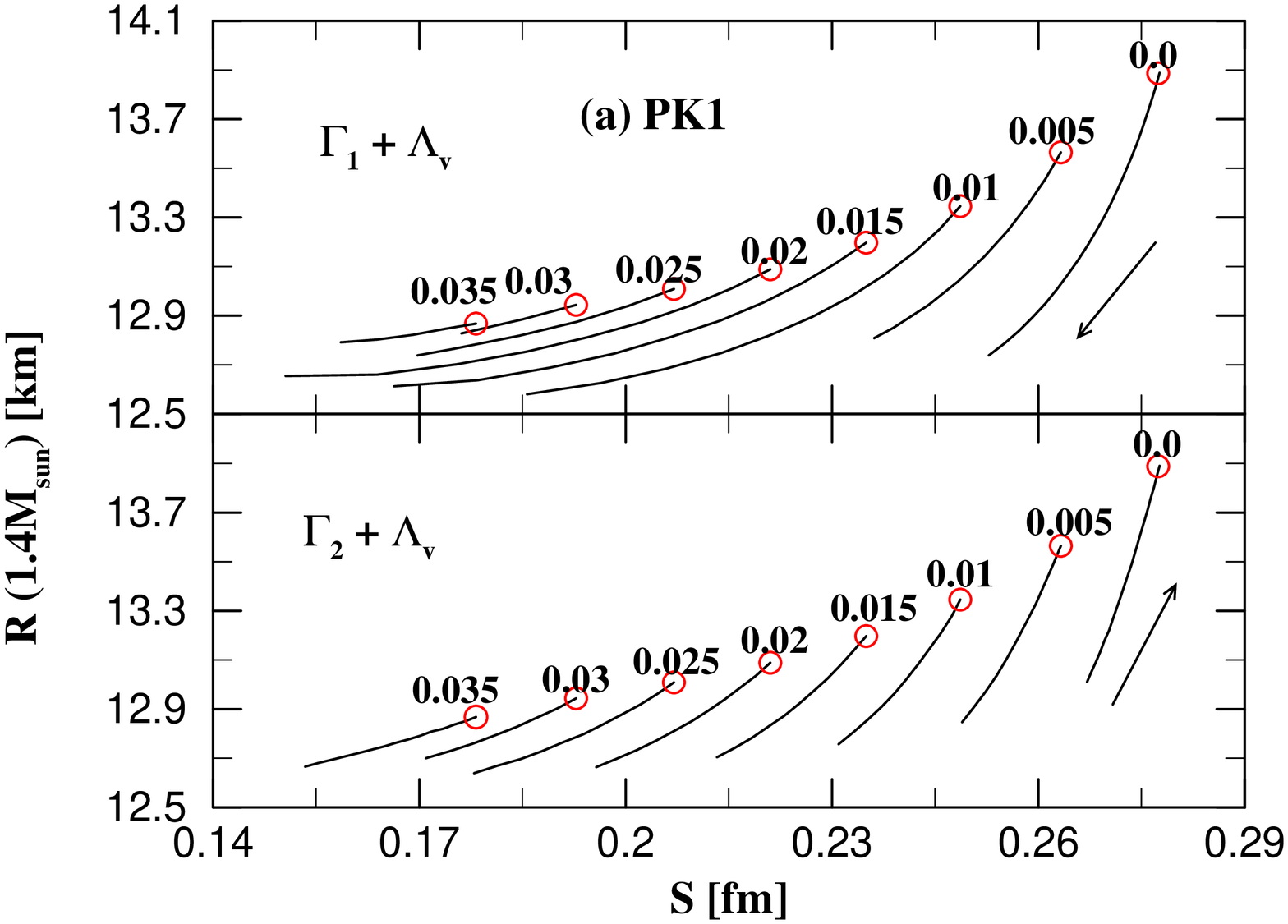}
\includegraphics[width=6cm,angle=-90]{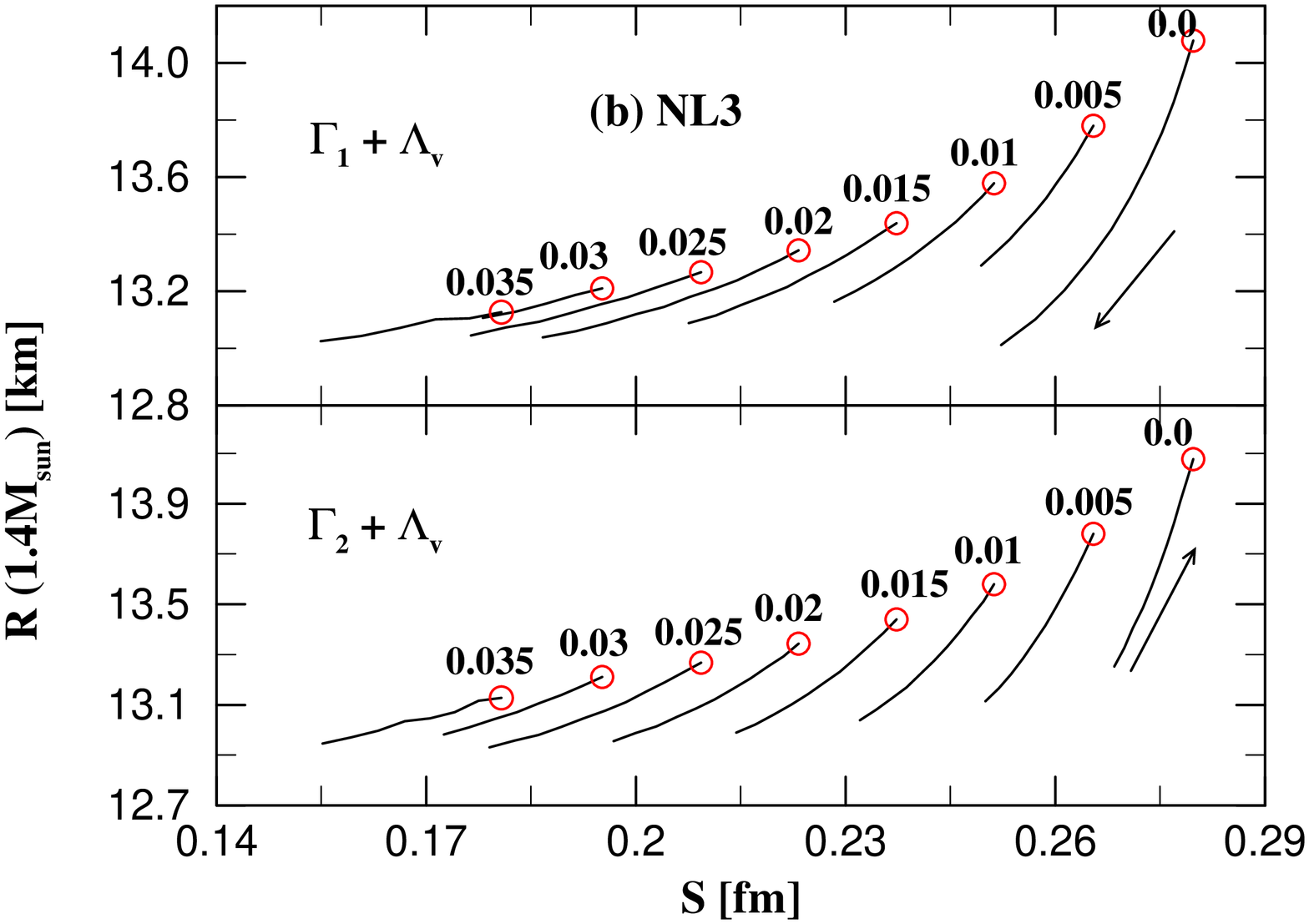}
\includegraphics[width=6cm,angle=-90]{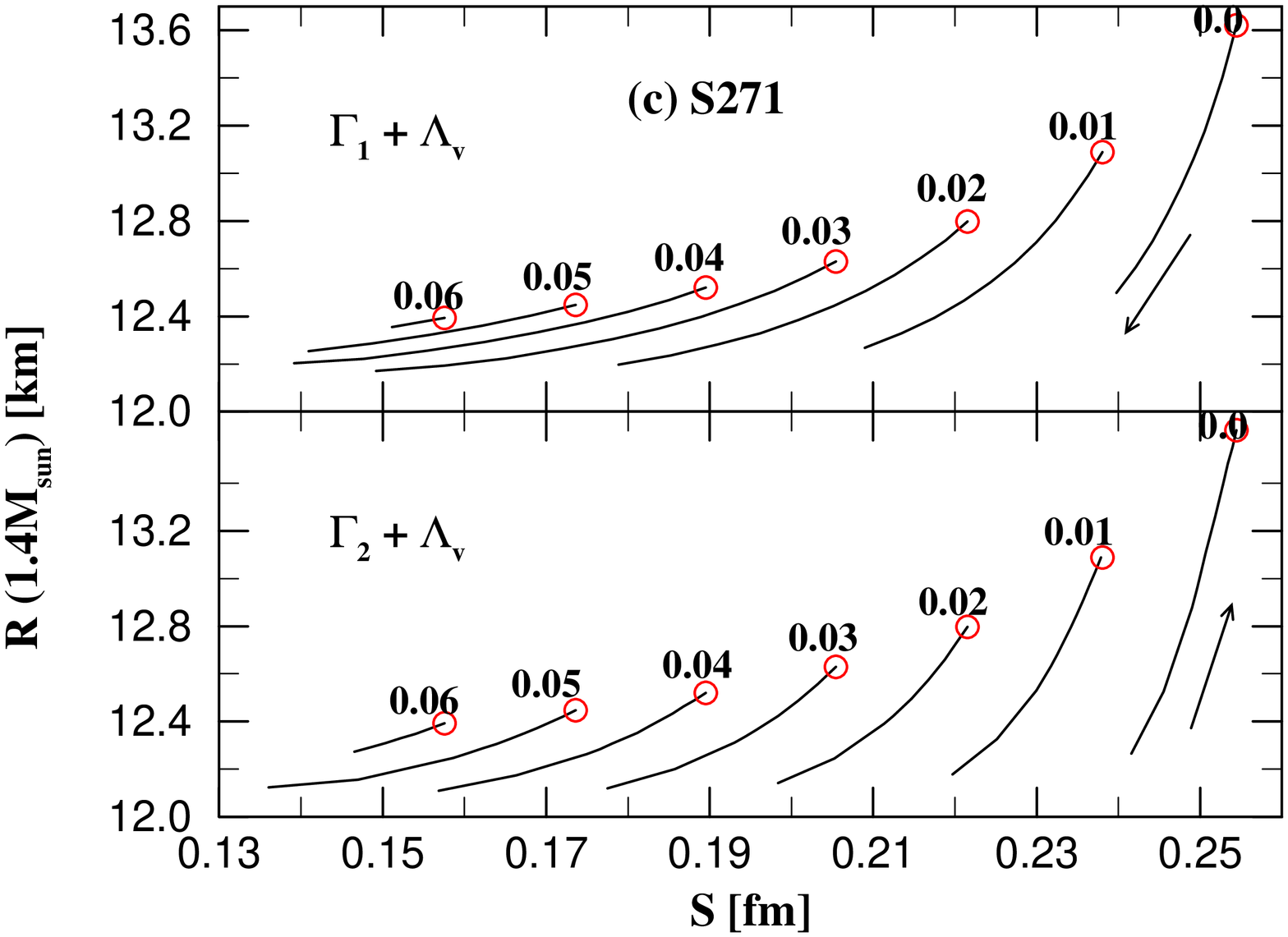}
\includegraphics[width=6cm,angle=-90]{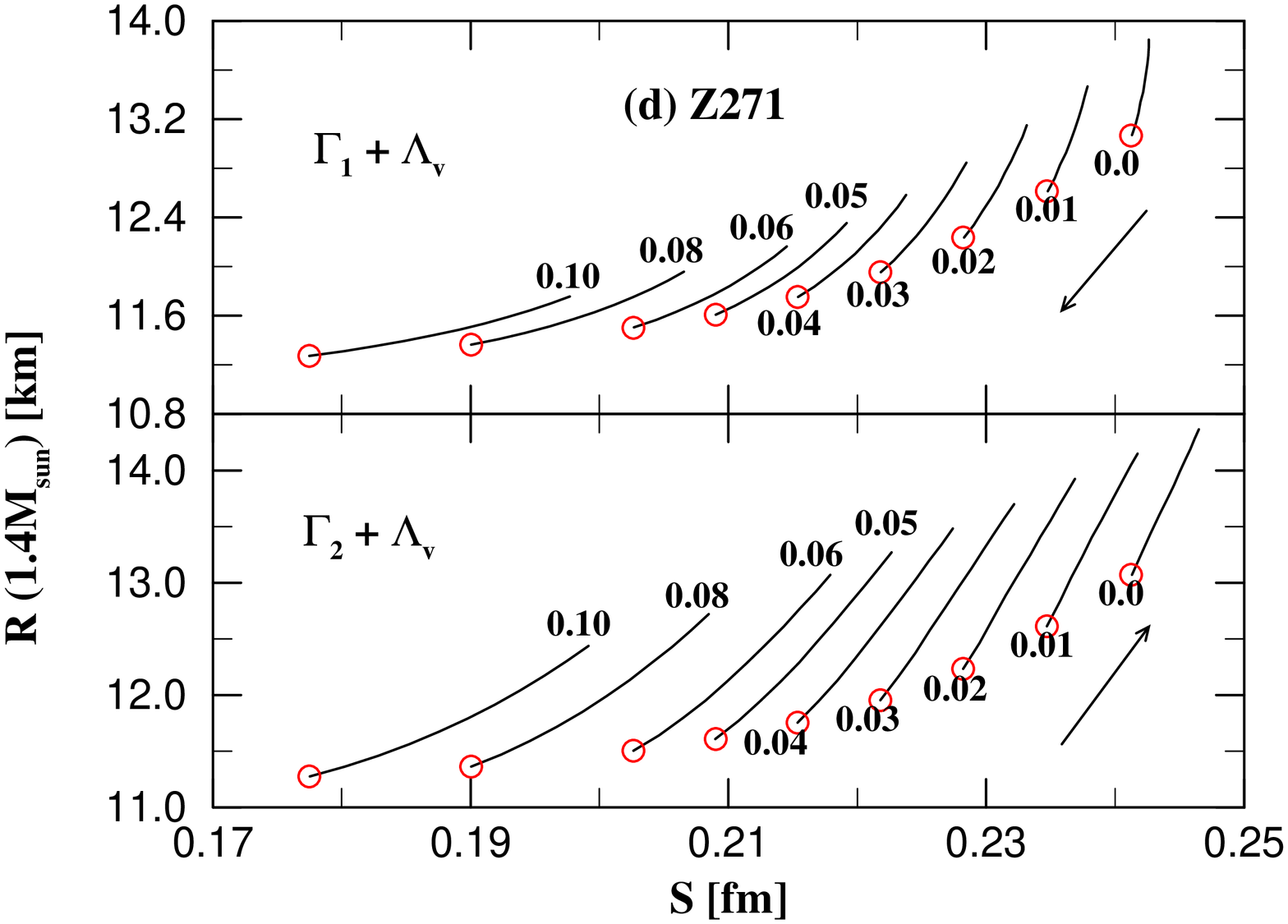}
\caption{(Color online) The radius, $R$ (in km), of a 1.4
solar-mass neutron star versus the neutron skin thickness, $S$ (in
fm), in $^{208}$Pb for the (a) PK1, (b) NL3, (c) S271, and (d)
Z271 effective interactions for various combinations of
$\Lambda_v$, $\Gamma_1$ and $\Gamma_2$. The upper panels are
associated with the $\Lambda_v$ and $\Gamma_1$, with $\Gamma_2 =
0$, combination: open circles denote values corresponding to
$\Gamma_1 = 0$ and the direction of the arrows indicates values of
increasing $\Gamma_{1}$. The lower panels are associated with the
$\Lambda_v$ and $\Gamma_2$, with $\Gamma_1 = 0$, combination: open
circles denote values corresponding to $\Gamma_2 = 0$ and the
direction of the arrows indicates values of increasing
$\Gamma_{2}$.} \label{fig:gamma_lambdav_all.eps}
\end{figure}

The effect of different combinations of $\Lambda_{v}$ with
$\Gamma_{1}$, with $\Gamma_{2} = 0$, and $\Lambda_{v}$ with
$\Gamma_{2}$,with $\Gamma_{1} = 0$, on the correlation between $S$
and $R$ for the (a) PK1, (b) NL3, (c) S271 and (d) Z271 effective
interactions is displayed in Fig.~\ref{fig:gamma_lambdav_all.eps}.
It is clear that the $\Lambda_v$ term, given by Eq.~(\ref{LHP}),
plays the most prominent role for decreasing the values of $R$ and
$S$, i.e. the $\Lambda_{v}$ term has the largest effect on
changing values of $R$ and $S$. In the upper left panel,
associated with the PK1 effective interaction, the minimum neutron
skin thickness in $^{208}$Pb is $S$ = 0.151~fm for the combination
$\Lambda_v = 0.02$, $\Gamma_1 = 0.8$, and $\Gamma_2 = 0.0$, while
the minimum radius of a 1.4 solar-mass neutron star is as low as
$R$ = 12.58~km when $\Lambda_v = 0.01$, $\Gamma_1 = 1.0$, and
$\Gamma_2 = 0.0$. The corresponding results for the NL3, S271, and
Z271 effective interactions are shown in the other panels in
Fig.~\ref{fig:gamma_lambdav_all.eps}. The lowest value of the
neutron skin thickness in $^{208}$Pb is $S$ = 0.136~fm for the
S271 effective interaction with $\Lambda_v = 0.05$, $\Gamma_1 =
0.0$, and $\Gamma_2 = -2.0$, and the smallest value of $R$ is
11.28~km for Z271 with $\Lambda_v = 0.1$, $\Gamma_1 = 0.0$, and
$\Gamma_2 = 0.0$.

\section{\label{sec:summary}Summary}

After introducing higher-order nucleon-sigma-rho coupling
corrections to existing RMF models, and calibrating the various
coupling constants with respect to the properties of nuclear
matter, we observe a new range of values for the neutron skin
thickness, $S$, in $^{208}$Pb, namely 0.16 $\sim$ 0.28~fm for the
PK1 and NL3 effective interactions and 0.11 $\sim$ 0.26~fm for the
S271 and Z271 effective interactions. This means that, on one
hand, for some particular effective interactions (S271 and Z271)
it is possible to obtain values of the neutron radius in the same
range as those for nonrelativistic mean field models, while on the
other hand, for other effective interactions (PK1 and NL3) the
neutron skin thickness is slightly extended compared to the
previous upper limit of 0.2~fm.

The addition of our new terms also has the effect of softening the
density dependence of the symmetry energy at high densities. We
also observe a linear relation between the neutron skin thickness
and the nuclear symmetry energy at the saturation point.

We have extrapolated our new relativistic mean field models from
normal to dense neutron matter so as to correlate the radius
of a 1.4 solar-mass neutron star with the neutron skin thickness in
$^{208}$Pb. We observe that the radius of a 1.4 solar-mass neutron
star is not uniquely constrained by a measurement of the neutron skin
thickness and is also model-dependent.

\begin{acknowledgments}

This work is partly supported by the Major State Basic Research
Development Program Under Contract Number G2000077407, the
National Natural Science Foundation of China under Grant No.
10025522 and 10221003, as well as the National Research Foundation
of South Africa under Grant No. 2054166. GCH also acknowledges
invaluable discussions with Chuck Horowitz, from Indiana
University in Bloomington, Indiana USA, which served as the
original catalyst for this project.

\end{acknowledgments}

\clearpage


\begin{table}[h]
\centering \caption{ Summary of current values for the neutron
skin thickness, $S$, in $^{208}$Pb} \label{tab:data}\btab{cccc}
\hline \hline

Probe &  $S$ (fm) &Error (fm)& Reference \\
\hline

$\pi^+$ and $\pi^-$ scattering & 0.0& 0.1  & \cite{Allar} \\
Proton scattering (650~MeV)  & 0.20&0.04 &\cite{staro}\\
Giant dipole resonance excitation & 0.19& 0.09  &\cite{kras}\\
Nucleon scattering (40 - 200~MeV)& 0.17 & &\cite{kara}\\
Proton-nucleus scattering (0.5-1.04 GeV)& 0.097 & 0.014  & \cite{clark}\\
Anti-protonic atoms & 0.15& 0.02  & \cite{Trz}\\
\hline \etab
\end{table}

\begin{table}[h]
\centering \caption{ Parameter sets associated with the original
PK1, NL3, S271, and Z271 RMF models. The nucleon and rho meson
masses are kept fixed at $M$ = 939~MeV for both protons, $p$, and
neutrons, $n$, and $m_\rho$ = 763~MeV for the NL3, S271, and Z271
effective interactions. For the PK1 effective interaction, M$_n$ =
939.5731~MeV, M$_p$ = 938.2796~MeV and $m_\rho$ = 763~MeV.}
\label{tab:para}\btab{cccccccc} \hline \hline RMF model &
$g_\sigma$ & $g_\omega$ & $g_2$~(fm$^{-1}$) & $g_{3}$ & $c_{3}$ &
$m_\sigma$~(MeV) &
$m_\omega$~(MeV) \\
\hline
PK1 & 10.3222&13.0131&-8.1688&-9.9976&55.636&514.0819&784.254\\
NL3 & 10.217&12.868&-10.431&-28.885&0&508.194&782.5\\
S271&9.006&10.806&-12.37&-17.323&0&505&783\\
Z271&7.031&8.406&-5.435&63.691&49.941&465&783\\
\hline \etab
\end{table}

\clearpage

\setlength{\tabcolsep}{2pt}
\begin{table}[h]
\centering \caption{Results for the PK1 effective interaction with
$\Gamma_2$ = 0. For each $\Lambda_v$, the range of $\Gamma_1$ is
given with the corresponding $g_\rho$ coupling constant. Also listed
are the binding energy per nucleon, $E/A$, the proton root mean square
radius, $R_p$, the neutron skin thickness, $S$, in $^{208}$Pb, and
the radius, $R$, of a 1.4 solar-mass neutron star.} \label{tab:PK11} \btab{c| c c c c c c }
\toprule[1pt]\toprule
\multirow{1}{0.7cm}{~$\Lambda_v$~}&\multicolumn{1}{c}{~$\Gamma_1$~}&\multicolumn{1}{c}{~~~~4$g_\rho^2$~~~~}
&\multicolumn{1}{c}{$E/A$~(MeV)}&\multicolumn{1}{c}{~$R_p$~(fm)}&\multicolumn{1}{c}{$S$~(fm)}
&\multicolumn{1}{c}{~~$R$~(km)~}\\
\hline
      & 0.00000  & 87.75942   & -7.85382  & 5.44398  & 0.26332  & 13.5650  \\
0.005 &   0.05000&  90.55426&  -7.84285&   5.44407&   0.26224& 13.4509  \\
\hline
      & 0.00000  & 93.54758   & -7.86507  & 5.44561  & 0.24868  & 13.3450  \\
0.01  & 0.10000  & 100.40040  & -7.84225  & 5.44608  & 0.24558  & 13.2510 \\
\hline
      & 0.00000  & 101.28410  & -7.87067  & 5.44764  & 0.23500  & 13.1970 \\
0.015 &   0.15000& 113.12450&  -7.83869&   5.44906&   0.22817  &13.0560\\
\hline
      & 0.00000  & 109.87232  & -7.87696  & 5.45031  & 0.22105  & 13.0880 \\
0.02      &  0.15000 & 123.96595&  -7.84521  & 5.45239 &  0.21255& 12.9596\\
\hline
      & 0.00000  & 120.07776  & -7.88201  & 5.45351  & 0.20705  & 13.0080  \\
0.025      & 0.20000  & 143.71214  & -7.83903  & 5.45747  & 0.19295  & 12.8760  \\
\hline
      & 0.00000  & 132.38804  & -7.88579  & 5.45726  & 0.19285  & 12.9440 \\
0.03   & 0.20000  & 161.69666  & -7.84229  & 5.46243  & 0.17613  & 12.8290  \\
\hline
        & 0.00000 & 147.51000    &  -7.88824 &  5.46158 &  0.17828 &        12.8950\\
0.035  &          0.20000         &   184.83000     &  -7.84363 &  5.46817 &  0.15861& 12.7910\\
 \hline \bottomrule\bottomrule \etab
\end{table}
\clearpage

\setlength{\tabcolsep}{2pt}
\begin{table}[h]
\centering \caption{Results for the PK1 effective interaction with
$\Gamma_1$ = 0. For each $\Lambda_v$, the range of $\Gamma_2$ is
given with the corresponding $g_\rho$ coupling constant. Also listed
are the binding energy per nucleon, $E/A$, the proton root mean square
radius, $R_p$, the neutron skin thickness, $S$, in $^{208}$Pb, and
the radius, $R$, of a 1.4 solar-mass neutron star.} \label{tab:PK12} \btab{c| c c c c c  c}
\toprule[1pt]\toprule
\multirow{1}{0.7cm}{~$\Lambda_v$~}&\multicolumn{1}{c}{~$\Gamma_2$~}&\multicolumn{1}{c}{~~~~4$g_\rho^2$~~~~}
&\multicolumn{1}{c}{$E/A$~(MeV)}&\multicolumn{1}{c}{~$R_p$~(fm)}&\multicolumn{1}{c}{$S$~(fm)}
&\multicolumn{1}{c}{~~$R$~(km)~}\\
\hline
      & 0.00000  & 87.75942   & -7.85382  & 5.44398  & 0.26332  & 13.5650  \\
0.005      &-0.10000  & 89.34030   & -7.84601  & 5.44358  & 0.26214  & 13.4870 \\
\hline
      & 0.00000  & 93.54758   & -7.86507  & 5.44561  & 0.24868  & 13.3450  \\
0.01      &-0.30000  & 99.48068   & -7.84046  & 5.44480  & 0.24426  & 13.1470  \\
\hline
      & 0.00000  & 101.28410  & -7.87067  & 5.44764  & 0.23500  & 13.1970 \\
0.015      &-0.30000  & 107.86900  & -7.84779  & 5.44734  & 0.22936  & 13.0240    \\
\hline
      &  0.00000  & 109.87232  & -7.87696  & 5.45031  & 0.22105  & 13.0880  \\
0.02      &-0.40000  & 120.51648  & -7.84641  & 5.45060  & 0.21207  & 12.8950   \\
\hline
      &  0.00000  & 120.07776  & -7.88201  & 5.45351  & 0.20705  & 13.0080  \\
0.025      & -0.50000  & 136.42240  & -7.84380  & 5.45487  & 0.19386  & 12.7980   \\
\hline
      &  0.00000  & 132.38804  & -7.88579  & 5.45726  & 0.19285  & 12.9440  \\
0.03      &-0.60000  & 157.20144  & -7.83931  & 5.46024  & 0.17446  & 12.7290\\
\hline
       & 0.00000 & 147.51000   &  -7.88824 &  5.46158 &  0.17828 &        12.8950\\
  0.035  &          -0.55000 &     175.83000 &    -7.84514  & 5.46548 &  0.15931 &   12.7150      \\
 \hline\bottomrule\bottomrule \etab
\end{table}
\clearpage


\setlength{\tabcolsep}{2pt}
\begin{table}[h]
\centering \caption{Results for the NL3 effective interaction with
$\Gamma_2$ = 0. For each $\Lambda_v$, the range of $\Gamma_1$
is given with the corresponding $g_\rho$ coupling constant. Also listed
are the binding energy per nucleon, $E/A$, the proton root mean square
radius, $R_p$, the neutron skin thickness, $S$, in $^{208}$Pb, and
the radius, $R$, of a 1.4 solar-mass neutron star.} \label{tab:NL31} \btab{c| c c c c  c c}
\toprule[1pt]\toprule
\multirow{1}{0.7cm}{~$\Lambda_v$~}&\multicolumn{1}{c}{~$\Gamma_1$~}&\multicolumn{1}{c}{~~~~4$g_\rho^2$~~~~}
&\multicolumn{1}{c}{$E/A$~(MeV)}&\multicolumn{1}{c}{~$R_p$~(fm)}&\multicolumn{1}{c}{$S$~(fm)}
&\multicolumn{1}{c}{~~$R$~(km)~}\\
\hline
      & 0.00000  & 84.86094   & -7.86252  & 5.46039 &  0.26546  & 13.7800\\
0.005      & 0.10000  & 90.36404   & -7.84141  & 5.46047 &  0.26303  & 13.6790 \\
\hline
      & 0.00000  & 90.85902   & -7.87088  & 5.46146 &  0.25128  & 13.5790\\
0.01      & 0.15000  & 100.68116  & -7.83939  & 5.46208 &  0.24585  & 13.4440 \\
\hline
      & 0.00000  & 97.85166   & -7.87783  & 5.46320  & 0.23729  & 13.4390\\
0.015      & 0.15000  & 109.28612  & -7.84707  & 5.46441  & 0.23016  & 13.3270 \\
\hline
      & 0.00000  & 105.92526  & -7.88395  & 5.46557  & 0.22333  & 13.3430\\
0.02      & 0.20000  & 124.63490  & -7.84268  & 5.46822  & 0.21123  & 13.2090 \\
\hline
      & 0.00000  & 115.51950  & -7.88873  & 5.46853  & 0.20935  & 13.2670 \\
0.025      & 0.20000  & 138.10950  & -7.84746  & 5.47226  & 0.19486  & 13.1530\\
\hline
      & 0.00000  & 127.01290  & -7.89232  & 5.47206  & 0.19521  & 13.2110\\
0.03      & 0.25000&  163.48180  &   -7.83926  & 0.17318   &5.47856 &13.0601\\
\hline
&       0.00000  &     141.00000   &   -7.89472  & 5.47619 &  0.18076  &   13.1630       \\
 0.035  &0.25000 & 187.42000 &  -7.84020    & 5.48457& 0.15495 &    13.0500      \\
  \hline\bottomrule\bottomrule
\etab
\end{table}
\clearpage


\setlength{\tabcolsep}{2pt}
\begin{table}[h]
\centering \caption{Results for the NL3 effective interaction with
$\Gamma_1$ = 0. For each $\Lambda_v$, the range of $\Gamma_2$ is
given with the corresponding $g_\rho$ coupling constant. Also listed
are the binding energy per nucleon, $E/A$, the proton root mean square
radius, $R_p$, the neutron skin thickness, $S$, in $^{208}$Pb, and
the radius, $R$, of a 1.4 solar-mass neutron star.} \label{tab:NL32} \btab{c| c c c c c  c}
\toprule[1pt]\toprule
\multirow{1}{0.7cm}{~$\Lambda_v$~}&\multicolumn{1}{c}{~$\Gamma_2$~}&\multicolumn{1}{c}{~~~~4$g_\rho^2$~~~~}
&\multicolumn{1}{c}{$E/A$~(MeV)}&\multicolumn{1}{c}{~$R_p$~(fm)}&\multicolumn{1}{c}{$S$~(fm)}
&\multicolumn{1}{c}{~~$R$~(km)~}\\
\hline
      &  0.00000  & 84.86094   & -7.86252  & 5.46039  & 0.26546  & 13.7800  \\
0.005      & -0.30000  & 89.60516   & -7.83969  & 5.45879  & 0.26154  & 13.5590  \\
\hline
      &  0.00000  & 90.85902   & -7.87088  & 5.46146  & 0.25128  & 13.5790 \\
0.01      &-0.40000  & 98.28740   & -7.84082  & 5.45993  & 0.24456  & 13.3320  \\
\hline
      &  0.00000  & 97.85166   & -7.87783  & 5.46320  & 0.23729  & 13.4390  \\
0.015      & -0.50000  & 108.86836  & -7.84069  & 5.46217  & 0.22699  & 13.1880  \\
\hline
      &  0.00000  & 105.92526  & -7.88395  & 5.46557  & 0.22333  & 13.3430  \\
0.02      & -0.60000  & 121.92576  & -7.83948  & 5.46551  & 0.20865  & 13.0860 \\
\hline
      &  0.00000  & 115.51950  & -7.88873  & 5.46853  & 0.20935  & 13.2670 \\
0.025      & -0.60000  & 134.79210  & -7.84435  & 5.46965  & 0.19256  & 13.0470 \\
\hline
      &  0.00000  & 127.01290  & -7.89232  & 5.47206  & 0.19521  & 13.2110  \\
0.03      & -0.70000  & 155.35130  & -7.83993  & 5.47508  & 0.17258  & 12.9820  \\
\hline
&       0.00000  &     141.00000   &   -7.89472 &   5.47619  &  0.18076  & 13.1630       \\
0.035   &    -0.65000   &    173.66000    &-7.84539 &   5.48046  & 0.15725  &  12.9770       \\
 \hline\bottomrule\bottomrule \etab
\end{table}
\clearpage

\setlength{\tabcolsep}{2pt}
\begin{table}[h]
\centering \caption{Results for the S271 effective interaction
with $\Gamma_2$ = 0. For each $\Lambda_v$, the range of $\Gamma_1$
is given with the corresponding $g_\rho$ coupling constant. Also
listed are the binding energy per nucleon, $E/A$, the proton root
mean square radius, $R_p$, the neutron skin thickness, $S$, in
$^{208}$Pb, and the radius, $R$, of a 1.4 solar-mass neutron star.}
\label{tab:S2711} \btab{c| c c c c c  c} \toprule[1pt]\toprule
\multirow{1}{0.7cm}{~$\Lambda_v$~}&\multicolumn{1}{c}{~$\Gamma_1$~}&\multicolumn{1}{c}{~~~~4$g_\rho^2$~~~~}
&\multicolumn{1}{c}{$E/A$~(MeV)}&\multicolumn{1}{c}{~$R_p$~(fm)}&\multicolumn{1}{c}{$S$~(fm)}
&\multicolumn{1}{c}{~~$R$~(km)~}\\
\hline
      &  0.00000  & 91.58490   & -7.94895  & 5.45986  & 0.23786  & 13.0880 \\
0.01      &  0.55000& 121.352251 &-7.84226  &     5.46037&0.22586   &12.5867\\
\hline
      &  0.00000  & 98.52548   & -7.96073  & 5.46078  & 0.22151  & 12.7970  \\
0.02  &0.65000& 141.943385 & -7.83858   &     5.46372   &  0.19862 & 12.3605  \\
\hline
      &  0.00000  & 106.66758  & -7.97100  & 5.46247  & 0.20543  & 12.6290 \\
0.03      & 0.70000  & 165.27674  & -7.84191  & 5.46869  & 0.17193  & 12.2640\\
\hline
      &  0.00000  & 116.33780  & -7.97978  & 5.46487  & 0.18950  &
      12.5210\\
0.04   & 0.75000  & 197.6836  & -7.84262    &    5.47557  &   0.14352& 12.3314\\
\hline
      &  0.00000  & 127.87086  & -7.98742  & 5.46795  & 0.17358  &
      12.4470\\
0.05
      & 0.80000  & 245.799698   & -7.83967   &  5.48465    & 0.11255&
        -\\
\hline \etab
\end{table}
\clearpage

\setlength{\tabcolsep}{2pt}
\begin{table}[h]
\centering \caption{Results for the Z271 effective interaction
with $\Gamma_2$ = 0. For each $\Lambda_v$, the range of $\Gamma_1$
is given with corresponding coupling constant $g_\rho$ between rho
meson and nucleon. Also listed are the binding energy per
nucleon, $E/A$, the proton root mean square radius, $R_p$, the neutron skin
thickness, $S$, in $^{208}$Pb, and the radius, $R$, of a 1.4 solar-mass
neutron star.} \label{tab:Z2711} \btab{c| c c c c  c c}
\toprule[1pt]\toprule
\multirow{1}{0.7cm}{~$\Lambda_v$~}&\multicolumn{1}{c}{~$\Gamma_1$~}&\multicolumn{1}{c}{~~~~4$g_\rho^2$~~~~}
&\multicolumn{1}{c}{$E/A$~(MeV)}&\multicolumn{1}{c}{~$R_p$~(fm)}&\multicolumn{1}{c}{$S$~(fm)}
&\multicolumn{1}{c}{~~$R$~(km)~}\\
\hline
      &  0.00000  & 92.582884  & -7.77718  & 5.45906  & 0.23478  & 12.6120 \\
0.01      &-0.50000  & 79.602084  & -7.83860  & 5.45937  & 0.23687  & 13.1190\\
\hline
      &  0.00000  & 95.023504  & -7.78256  & 5.45913  & 0.22827  & 12.2350 \\
0.02      &-0.50000  & 81.396484  & -7.84313  & 5.45928  & 0.23145  & 12.7610 \\
\hline
      &  0.00000  & 97.614400  & -7.78766  & 5.45931  & 0.22182  & 11.9530 \\
0.03      & -0.50000  & 83.283876  & -7.84748  & 5.45928  & 0.22607  & 12.4490\\
\hline
      &  0.00000  & 100.360324 & -7.79255  & 5.45960  & 0.21542  & 11.7530  \\
0.04      & -0.40000  & 88.021924  & -7.84028  & 5.45937  & 0.21984  & 12.1030 \\
\hline
      &  0.00000  & 103.22560  & -7.79745  & 5.46002  & 0.20901  & 11.6100  \\
0.05      &-0.40000  & 90.212004  & -7.84461  & 5.45960  & 0.21430  & 11.9140\\
\hline
      &  0.00000  & 106.29610  & -7.80200  & 5.46053  & 0.20267  & 11.5040  \\
0.06      &-0.40000  & 92.544400  & -7.84865  & 5.45990  & 0.20881  & 11.7690  \\
\hline
      &  0.00000  & 112.954384  & -7.81084  & 5.46188  & 0.19004  & 11.3630 \\
0.08      &-0.30000  & 101.123136  & -7.84530  & 5.46099  & 0.19613  & 11.5130 \\
\hline
      &  0.00000  & 120.56040   & -7.81876  & 5.46361  & 0.17751  & 11.2750  \\
0.10      & -0.20000  & 111.386916  & -7.84152  & 5.46268  & 0.18260  & 11.3530  \\
\hline\bottomrule\bottomrule \etab
\end{table}
\clearpage

\clearpage

\clearpage

\clearpage

\clearpage

\clearpage

\clearpage

\clearpage

\clearpage

\clearpage

\newpage
\newpage

\newpage

\newpage

\newpage
\clearpage


\end{document}